\documentclass{aa}
\usepackage{graphicx}
\usepackage{txfonts}
\usepackage{textcomp, gensymb}
\usepackage{lscape}
\usepackage{xcolor}
\usepackage[version=4]{mhchem} 
\usepackage[colorlinks = true,
            linkcolor = blue,
            urlcolor  = blue,
            citecolor = blue,
            anchorcolor = blue, unicode]{hyperref}

\makeatletter
\renewcommand*\aa@pageof{, page \thepage{} of \pageref*{LastPage}}
\makeatother

\begin{document}

   \title{Mid-infrared spectra of T Tauri disks: modeling the effects of a small inner cavity on \ce{CO2} and \ce{H2O} emission}


   \author{Marissa Vlasblom
          \inst{1}
          \and
          Ewine F. van Dishoeck\inst{1,2}
          \and
          Beno\^it Tabone \inst{3}
          \and
          Simon Bruderer \inst{2} } 

   \institute{Leiden Observatory, Leiden University, 2300 RA Leiden, Netherlands\\
              \email{vlasblom@strw.leidenuniv.nl}
              \and
              Max-Planck-Institut f\"ur Extraterrestrische Physik, Giessenbachstrasse 1, 85748 Garching, Germany
              \and
              Universit\'e Paris-Saclay, CNRS, Institut d’Astrophysique Spatiale, 91405 Orsay, France
             }

   \date{Received 11 October 2023; accepted 17 November 2023}

   \titlerunning{Modeling the effects of a small inner cavity on the \ce{CO2} and \ce{H2O} spectra}
   \authorrunning{Vlasblom et al.}

\abstract
    {The inner few AU of disks around young stars, where terrestrial planets are thought to form, are best probed in the infrared. The {\it James Webb} Space Telescope (JWST) is now starting to characterize the chemistry of these regions in unprecedented detail, building on earlier results of the {\it Spitzer} Space Telescope that planet-forming zone of disks contain a rich chemistry. One peculiar subset of sources characterized by {\it Spitzer} are the so-called ``\ce{CO2}-only sources'', in which only a strong 15 $\mu$m \ce{CO2} feature was detected in the spectrum.}
    {One scenario that could explain the weak or even non-detections of molecular emission from \ce{H2O} is the presence of a small, inner cavity in the disk. If this cavity were to extend past the \ce{H2O} snowline, but not past the \ce{CO2} snowline, this could strongly suppress the \ce{H2O} line flux w.r.t. that of \ce{CO2}. In this work, we aim to test the validity of this statement. }
    {Using the thermo-chemical code Dust And LInes (DALI), we created a grid of T Tauri disk models with an inner cavity, meaning we fully depleted the inner region of the disk in gas and dust starting from the dust sublimation radius and ranging until a certain cavity radius. Cavity radii varying in size from 0.1 to 10 AU are explored in this work. We extended this analysis to test the influence of cooling through \ce{H2O} ro-vibrational lines and the luminosity of the central star on the \ce{CO2 /H2O} flux ratio.  }
    {We present the evolution of the \ce{CO2} and \ce{H2O} spectra of a disk with inner cavity size. The line fluxes show an initial increase as a result of increasing emitting area, followed by a sharp decrease. As such, when a large-enough cavity is introduced, a spectrum that was initially dominated by \ce{H2O} lines can become \ce{CO2}-dominated instead. However, the cavity size needed for this is around 4-5 AU, exceeding the nominal position of the \ce{CO2} snowline in a full disk, which is located at 2 AU in our fiducial, $L_* = 1.4 \, L_\odot$ model. The cause of this is most likely the alteration of the thermal structure by the cavity, which pushes the snowlines outward. In contrast, our models show that global temperature fluctuations, for example due to changes in stellar luminosity, impact the fluxes of \ce{H2O} and \ce{CO2} roughly equally, thus not impacting their ratio much. Alternative explanations for bright \ce{CO2} emission are also briefly discussed.}
    {Our modeling work shows that it is possible for the presence of a small inner cavity to explain strong \ce{CO2} emission in a spectrum. However, the cavity needed to do so is larger than what was initially expected. As such, this scenario will be easier to test with sufficiently high angular resolution (millimeter) observations. }  
 


   \maketitle
  
%

\section{Introduction}
\label{sec:intro}
The formation of terrestrial planets is thought to occur mostly within the inner few AU of protoplanetary disks around young stars \citep{morbidelli2012}. To understand the composition of these planets and their atmospheres, investigating the chemistry of their birth environment is crucial. This environment is certainly not static, as the inner disk hosts strong variations in temperature, density and UV irradiation with position. Additionally, the presence of substructures and drift of icy grains can further alter the composition of the inner disk. Tracing the inner disk structure can be done though infrared (IR) spectroscopy. Previous surveys conducted with, for example, the \textit{Spitzer} Space Telescope, have found most disks to contain a rich chemistry, with the molecules \ce{H2O, CO2, CO, HCN and C2H2} commonly detected in the 10-20 $\mu$m range \citep{carr2008, pontoppidan2010, salyk2011, pontoppidan2014}. Now, the \textit{James Webb} Space Telescope \citep[JWST;][]{gardner2023, rigby2023} provides even better insights into the complexities of these warm, inner regions. The higher sensitivity and higher spectral resolution of JWST compared with \textit{Spitzer} boosts the line-to-continuum ratio, allowing it to detect much weaker emission from less abundant, minor species and for even better quantitative interpretation of molecular emission, due to these bands being much less blended than they were when observed with \textit{Spitzer}. \\
\begin{figure*}[t]
    \makebox[\textwidth][c]{\includegraphics[scale=0.55]{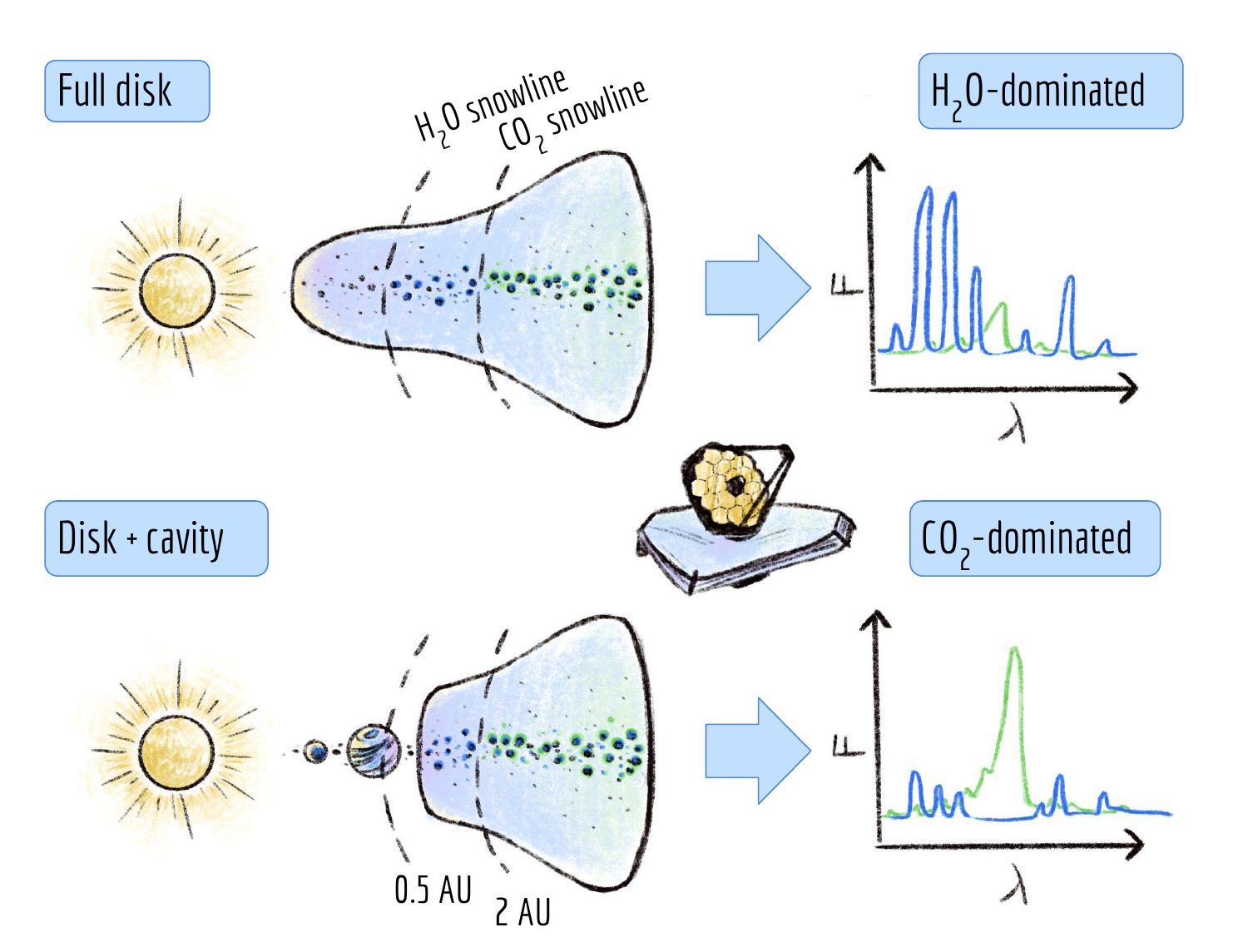}}
    \caption{Cartoon illustrating the scenario that is tested in this work. On the top, a full disk is shown, leading to an observed spectrum with bright \ce{H2O} lines (illustrated in blue) and a relatively weaker 15 $\mu$m \ce{CO2} feature (illustrated in green). On the bottom, a disk with a small inner cavity is illustrated. This could, for example, be carved by a companion. The cavity extends beyond the \ce{H2O} snowline but not beyond the \ce{CO2} snowline, leading to an observed spectrum that shows a bright \ce{CO2} feature and relatively weak \ce{H2O} lines. The values given for the locations of the \ce{H2O} and \ce{CO2} snowlines are the midplane snowline locations of our fiducial, $L_* = 1.4 L_\odot$ full disk model based on AS 209. }
    \label{fig:cartoon}
\end{figure*}
\newline
In the mid-IR wavelength range probed by the Mid-InfraRed Instrument \citep[MIRI;][]{wright2015, wright2023, rieke2015}, emission lines from \ce{H2O} and \ce{CO2} are often prominently detected \citep{kamp2023, vandishoeck2023}, as was also already the case with \textit{Spitzer}. To interpret these findings, many sophisticated models of the inner disks of T Tauri stars were developed over the years \citep[e.g.][]{agundez2008, woitke2009, walsh2010, najita2011, bruderer2012}, with several of them being specifically focused on the formation and evolution of \ce{H2O} \citep{glassgold2009, bethell2009, meijerink2009, adamkovics2014, du2014}. Recent modeling results have shown that the \ce{H2O} and \ce{CO2} abundances and their emission are sensitive to the gas temperature \citep[see, e.g.][]{walsh2015, woitke2018, anderson2021}, making their emission and their relative flux levels interesting tracers of the inner disk's physical and chemical conditions. Namely, \ce{H2O} and \ce{CO2} share a common pathway in their gas-phase formation. Both molecules are formed from OH, but which of the two is formed most is determined by the gas temperature \citep{charnley1997, vandishoeck2013, walsh2015}. At temperatures between 100 and 250 K, OH reacts with CO to form \ce{CO2} via the reaction \ce{OH + CO -> CO2 + H}. Above a temperature of roughly 300 K, the energy barrier of the reaction \ce{OH + H2 -> H2O + H} at 1740 K \citep{baulch1992} can be overcome, and due to \ce{H2} being more abundant than \ce{CO}, the vast majority of the \ce{OH} will be turned into \ce{H2O} instead of \ce{CO2}. \\
\newline
However, in some sources observed with \textit{Spitzer}, only the 15 $\mu$m band from \ce{CO2} was detected, whereas \ce{H2O} emission seemed absent in these observations \citep{pontoppidan2010}. These sources became known as the ``\ce{CO2}-only sources''. One of these sources, GW Lup, has now been observed with JWST-MIRI as a part of the MIRI mid-INfrared Disk Survey (MINDS) \citep{grant2023, kamp2023}. While more molecules than just \ce{CO2} were detected, thanks to the higher sensitivity and resolving power of JWST compared to \textit{Spitzer}, the spectrum of GW Lup still stands out from typical IR T Tauri disk spectra in that its \ce{CO2} line flux is especially strong. Its \ce{CO2} flux is much stronger than the surrounding \ce{H2O} lines at 15 $\mu$m, which is not something that is commonly seen in these sources. As such, even emission from \ce{^13CO2} was detected in this disk, which had not been achieved before. Also the disk's \ce{CO2 /H2O} column density ratio, estimated with 0D local thermodynamic equilibrium (LTE) slab models, is found to be particularly high compared with other sources, at a value of approximately 0.7 \citep{grant2023}. \\ 
\newline
A possible scenario to explain the presence of such a large \ce{CO2 /H2O} flux ratio was put forth: the presence of a small (few AU) cavity in the inner disk. Such a cavity in the millimeter dust continuum could, for example, be carved by a companion and would unlikely be resolved with the Atacama Large Millimeter/sub-millimeter Array (ALMA). If that cavity would extend beyond the \ce{H2O} snowline, it could suppress the \ce{H2O} emission. Additionally, it would also prevent any remaining \ce{H2O} within the cavity to be efficiently replenished by radial drift. If the cavity does not extend further than the \ce{CO2} snowline, however, this would not happen to the \ce{CO2} emission. If the \ce{H2O} emission can be suppressed  with respect to the \ce{CO2} emission in this manner, it could lead to the source's spectrum exhibiting a large \ce{CO2 /H2O} flux ratio, like is seen in the \ce{CO2}-only sources. This scenario is illustrated in Fig. \ref{fig:cartoon}. In this work, we set out to test this by modeling the effects of a small cavity on the \ce{H2O} and \ce{CO2} spectra of a T Tauri disk. However, there are also other scenarios that should be considered, which we will mention here and discuss in more detail in Sect. \ref{sec:discussion}.\\
\newline
First, the presence of a dust trap with only a small perturbation on the gas surface density, rather than a small, deep cavity, between the \ce{H2O} and \ce{CO2} snowlines could have a similar effect on the \ce{H2O} and \ce{CO2} line fluxes. A dust trap at this location could halt the drift of icy pebbles, preventing the sublimation of \ce{H2O} ice but not that of \ce{CO2} ice, allowing the \ce{CO2} gas to be continuously replenished while the \ce{H2O} gas is depleted. This could also lead to the spectrum exhibiting a large \ce{CO2 /H2O} flux ratio. \\
\newline
Second, since the gas temperature in a disk is largely set by the stellar luminosity, it would not be surprising for lower-luminosity stars to be more \ce{CO2}-rich, as the \ce{CO2} and \ce{H2O} abundances are sensitive to the gas temperature. In fact, low-mass stars have been shown, both by observations and models, to be richer in carbon-bearing species than their higher-mass counterparts \citep[e.g.][]{pascucci2013, walsh2015}. However, very-low-mass stars often show strong emission from hydrocarbons such as \ce{C2H2} (see, e.g. \citealt{tabone2023}, Arabhavi et al., submitted for recent examples with JWST), and this is not seen in these \ce{CO2}-only sources. Still, the \ce{CO2}-only class of sources could represent an in-between case, between the cold, very-low-mass stars rich in carbon-bearing species and the hotter, higher-mass stars rich in oxygen-bearing species. \\
\newline
Finally, the amount of small grains in the disk atmosphere can also impact molecular line fluxes. Since the gas-to-dust ratio has a strong impact on the line strengths \citep[see, e.g.][]{meijerink2009, woitke2018, greenwood2019}, the typically weak \ce{H2O} fluxes in the \ce{CO2}-only sources could point to these sources having a larger amount of dust in the upper layers, and thus a smaller gas-to-dust ratio. This may also impact the relative \ce{CO2 /H2O} flux ratios. After all, the smaller gas-to-dust ratio will lead to the disk atmosphere being cooler, which will promote the gas-phase formation of \ce{CO2} over \ce{H2O}. This effect can be tentatively seen in \citet{bosman2022b}.\\
\newline
For this work, we make use of the thermo-chemical code Dust And LInes \citep[DALI;][]{bruderer2012, bruderer2013}, which has previously been used to model \ce{CO2} mid-IR emission by \citet{bosman2017}. We include the most recent updates to the code as described in \citet{bosman2022a, bosman2022b}. The work by these authors introduces \ce{H2O} UV-shielding \citep{bethell2009} and so-called `(photo-)chemical heating' to the code. The latter refers to the process in which excess energy released by the photodissociation of molecules and the subsequent chemical reactions involving the dissociation products is used to heat the gas \citep{glassgold2015}. The work by \citet{bosman2022a, bosman2022b} finds both of these processes to be especially important in matching modeled spectra to typical \textit{Spitzer} observations, where there had previously been some difficulty to do so \citep[see][]{woitke2018, anderson2021}. In particular, models previously had a tendency to greatly overproduce the \ce{CO2} flux w.r.t the \ce{H2O} flux, leading to much higher \ce{CO2 /H2O} flux ratios than most sources show. The inclusion of \ce{H2O} UV-shielding and extra chemical heating were shown by \citet{bosman2022b} to help mitigate this problem. The reasoning behind this relies again on the common gas-phase formation pathway of \ce{CO2} and \ce{H2O}. Namely, the extra chemical heating increases the temperature in the upper layers of the disk, and thus stimulates the \ce{H2O} production in this region. As the \ce{H2O} becomes more abundant in the disk, eventually it will reach the column densities required for it to start UV-shielding \citep{bethell2009}. This will further drive the balance from \ce{CO2} towards \ce{H2O}, even at a slightly lower temperature. The \ce{H2O} is then photodissociated at a much lower rate and this also reduces the production rate of \ce{OH}, which quenches the formation of \ce{CO2}. Moreover, \ce{CO2} does not benefit from self-shielding at these column densities, so its photodissociation rate will be higher. So, the inclusion of both of these effects boosts the \ce{H2O} production in the disk and allows the models to closer match the fact that \ce{H2O} emission is typically stronger than \ce{CO2} emission in observed mid-IR spectra. Of course, this goes in the opposite direction to what we are interested in (sources with \ce{H2O} emission fainter than \ce{CO2}), but these effects are still important to include to better match what is generally observed in mid-IR spectra, rather than only our specific subset of \ce{CO2}-only sources.  \\
\newline
The structure of this work is as follows: we describe our modeling setup in Sect. \ref{sec:methods} and the results in Sect. \ref{sec:results}. The implications of our modeling work are discussed in Sect. \ref{sec:discussion} and our main conclusions are summarized in Sect. \ref{sec:conclusions}. 
\section{Methods}
\label{sec:methods}
In this work, we present 5 grids of thermo-chemical models made with DALI which have small inner cavities. The fiducial grid, which is presented in the main body of this work, consists of models that are based on those presented in \citet{bosman2022a, bosman2022b}, which in turn are based on the model of the disk around the pre-main-sequence K5 star AS 209 presented in \citet{zhang2021}. 
All grids contain models with inner cavities ranging in size from 0.1 to 10 AU, as well as a full disk model without a cavity. The importance of \ce{H2O} ro-vibrational cooling as well as input stellar luminosity are tested with additional grids which are presented in Appendices \ref{sec:app_rovib} and \ref{sec:app_lum}. We summarize the model parameters of all grids in Table \ref{tab:params} and provide the main details of our setup below.\\
\begin{table}[t]
    \caption{Summary of all model parameters.} 
    \centering
    \begin{tabular}{l c c}
        \hline
        \hline
         Parameter          & Symbol            & Value  \\
        \hline
        Stel. luminosity          & $L_*$             & [0.2, 0.4, 0.8, \textbf{1.4}] $L_\odot$\\
        Eff. temperature     & $T_{\rm eff}$     & [3500, 3750, \\
                            &                   & 4000, \textbf{4300}] K \\
        FUV luminosity & $\log L_{\rm FUV}/L_\odot$     & [-3.45, -2.92, \\
                        &               & -2,62, \textbf{-2.12}] \\
        &&\\
        Subl. radius  & $R_{\rm subl}$    & [0.03, 0.04, \\
                        &                   & 0.06, \textbf{0.08}] AU    \\  
        Cavity radius       & $R_{\rm cav}$     & [$R_{\rm subl}$, 0.1, 0.3, \\
                            &                   &  0.5, 0.8, 1, 2, 3, \\
                            &                   &   4, 5, 7, 10] AU \\
        Outer radius        & $R_{\rm out}$     & 100 AU \\
        Gas depl. factor    & $\delta_{\rm gas}$    & 10$^{-20}$ \\
        Dust depl. factor   & $\delta_{\rm dust}$   & 10$^{-20}$ \\
        Char. radius   & $R_{\rm c}$         & 46 AU \\
        Char. gas surf. dens.   & $\Sigma_{\rm c}$    & 21.32 g cm$^{-2}$ \\
        Power-law index     & $\gamma$          & 0.9 \\
        Char. scale height  & $h_{\rm c}$             & 0.08 \\
        Flaring index       & $\psi$            & 0.11 \\
        Large dust mass frac. & $f_\ell$             & 0.999 \\ 
        Large dust settl. par.              & $\chi$            & 0.2 \\
        &&\\
        C abundance         & C/H          &$1.35\cdot10^{-4}$\\
        O abundance         & O/H          &$2.88\cdot10^{-4}$\\
        \hline
    \end{tabular}
    \tablefoot{Indicated in boldface are the parameters associated with the fiducial grid based on the AS 209 models by \citet{bosman2022a, bosman2022b}.}
    \label{tab:params}
\end{table}
\newline
The code DALI generates the physical-chemical models in three main steps \citep[see][]{bruderer2012, bruderer2013}. In the first step, the local radiation field and dust temperature are determined at all locations in the disk by solving a dust radiative transfer calculation with a 2D Monte Carlo method. In the next step, the chemical abundances are calculated, assuming steady state chemistry and using an initial guess for the gas temperature. These abundances are then used as input for the non-LTE excitation calculation for several specified atoms and molecules that make up the main coolants of the model. The code then solves for the thermal balance to obtain an improved estimate for the gas temperature. Since the previously calculated quantities -- the chemical abundances and the atomic/molecular excitation -- are both dependent on the gas temperature, this process is repeated in an iterative sequence until a convergent solution is reached. Once convergence is achieved, a raytracing tool can be used in a third step to obtain line fluxes, spectra and spectral image cubes. \\
\newline
The radial structure of the gas and dust of a full, smooth disk (with no cavity or gap present) is parameterized as follows
\begin{align}
    \Sigma_{\rm gas} (R) = \Sigma_{\rm c} \left( \frac{R}{R_{\rm c}} \right)^{-\gamma} \exp \left[ - \left( \frac{R}{R_{\rm c}}\right)^{2-\gamma} \right],
    \label{eq:dens_structure}
\end{align}
based on a self-similar solution to a viscously evolving disk \citep{lyndenbell1974, hartmann1998}. Here, $\Sigma_{\rm gas}$ is the gas surface density, $\Sigma_{\rm c}$ is the gas surface density at the characteristic radius $R_{\rm c}$ and $\gamma$ is the power-law index. The vertical scale height of the gas is given by
\begin{align}
    h = h_{\rm c} \left( \frac{R}{R_{\rm c}} \right)^\psi,
\end{align}
where $h_{\rm c}$ is the characteristic scale height at the characteristic radius and $\psi$ is the flaring index of the disk. The dust is separated into two populations by their grain size: small grains have sizes between 5 nm and 1 $\mu$m and large grains have sizes between 5 nm and 1 mm, with both populations following an MRN distribution \citep{mathis1977}. The small grains are well coupled to the gas, and they follow the same radial and vertical distribution \citep{miotello2016}. Poly-cyclic aromatic hydrocarbons (PAHs) are not considered in this work. The large grains have a mass fraction determined by $f_\ell$ and have a similar vertical distribution with a reduced scale height of $\chi  h_{\rm c}$.  From the four models presented in \citet{bosman2022a, bosman2022b}, we base all of our grids on the thin ($h_c$ = 0.08) model with the largest value for the large dust fraction ($f_\ell$ = 0.999), meaning that almost all of the dust is large and has thus settled to the midplane, which gives these models a gas-to-dust ratio of 10$^5$ in the upper layers of the disk. \\
\newline
The chemical network used for this work is the same as the one used in \citet{bosman2022a, bosman2022b}. This network builds on the standard DALI chemical network, based on UMIST06 \citep{woodall2007}, by including \ce{H2O} UV-shielding and more efficient \ce{H2} formation at high temperatures. Specifically, several \ce{H2} formation routes via three-body reactions have been included \citep[see Appendix A in][]{bosman2022a} and the chemisorption binding energy of H has been increased from 10000 K to 30000 K, which allows \ce{H2} to form on grains at dust temperatures between 300 and 900 K \citep{cazaux2002, cazaux2004, wakelam2017, thi2020}. The binding energies for \ce{H2O} and \ce{CO2} are 4820 K and 2690 K respectively, following \citet{sandford1993, aikawa1997}. Isotope chemistry, like implemented in \citet{miotello2014, visser2018}, is not included in this network. The \ce{H2O}, \ce{^12CO2} and \ce{^13CO2} spectra are calculated using DALI's ``fast line ray-tracer'' \citep[][Appendix B]{bosman2017}.  For this procedure, we assume that the disk is located at a distance of 121 pc (the distance to AS 209, \citealt{gaia2018}) and has a face-on orientation. Since \ce{^13C} is not included in the chemical network, the \ce{^13CO2} spectrum is derived by scaling its abundance with \ce{^12CO2} according to the ISM \ce{^12C /^13C} ratio of 77 \citep{wilson1994}. All spectra are then convolved to a resolving power of $R$ = 3000, comparable to that of MIRI in this wavelength range. We focus our analysis of the spectra mainly on the 13-17 $\mu$m wavelength range, since the lines in this region trace similar regions within the disk. The molecular data file for \ce{H2O} is taken from the Leiden Atomic and Molecular Database, in which levels with energies up to 7200 K are included \citep{tennyson2001}, line transitions come from the BT2 list \citep{barber2006} and collisional rate coefficients are from \citet{faure2008}. The molecular data for \ce{CO2} come from \citet{bosman2017}, for which the energy levels, line positions and line strengths are taken from the HITRAN database \citep{rothman2013}, with collisional rate coefficients based on \citet{allen1980, nevdakh2003, jacobs1975}.\\
\newline
As stated previously, our fiducial grid consists of models that are based on the AS 209 model presented in \citet{zhang2021}. Hence, we use the stellar input spectrum from this paper and thus these models have a central source with spectral type K5, an effective temperature of 4300 K and a stellar luminosity of 1.4 $L_\odot$.  
The grids of models presented in this work contain a small, inner cavity. In this context, we use the word ``cavity'' to mean a gap in the disk that starts at the sublimation radius and extends until a certain specified radius, referred to as the ``cavity size''. Thus, in our use of the word, a disk with a cavity has no inner disk present. Our grids of models contain cavity sizes $R_{\rm cav}$ between 0.1 and 10 AU. The cavity itself is devoid of gas and dust. Each grid also contains a full disk model, for which the ``cavity size'' is equal to the sublimation radius, which has a value of 0.08 AU for our fiducial grid. 
The outer radius, $R_{\rm out}$, of all models is set to 100 AU. \\ 
\newline
Since the mid-IR emission of \ce{H2O} and \ce{CO2} that we model in this work traces the innermost regions of the disk, it is very important for our models to have sufficient resolution at the cavity wall. This is especially true for our models with larger cavity sizes, to prevent that all emission traces only a single or a few cells. As such, we have slightly modified how the grid is set up from $R_{\rm cav}$ to $R_{\rm out}$. In the normal DALI setup \citep[see][for details]{bruderer2013}, this region contains a single, logarithmic grid. We cover this area with two grids instead, a logarithmic grid that covers the cavity wall, and a linear grid that covers the rest of the disk. This allows us to get the very high resolution needed at the edge of the cavity. The transition point between the two grids is chosen to ensure a smooth transition in grid cell size, and lies well beyond the relevant emitting regions. \\
\begin{figure*}[t]
    \makebox[\textwidth][c]{\includegraphics[scale=0.45]{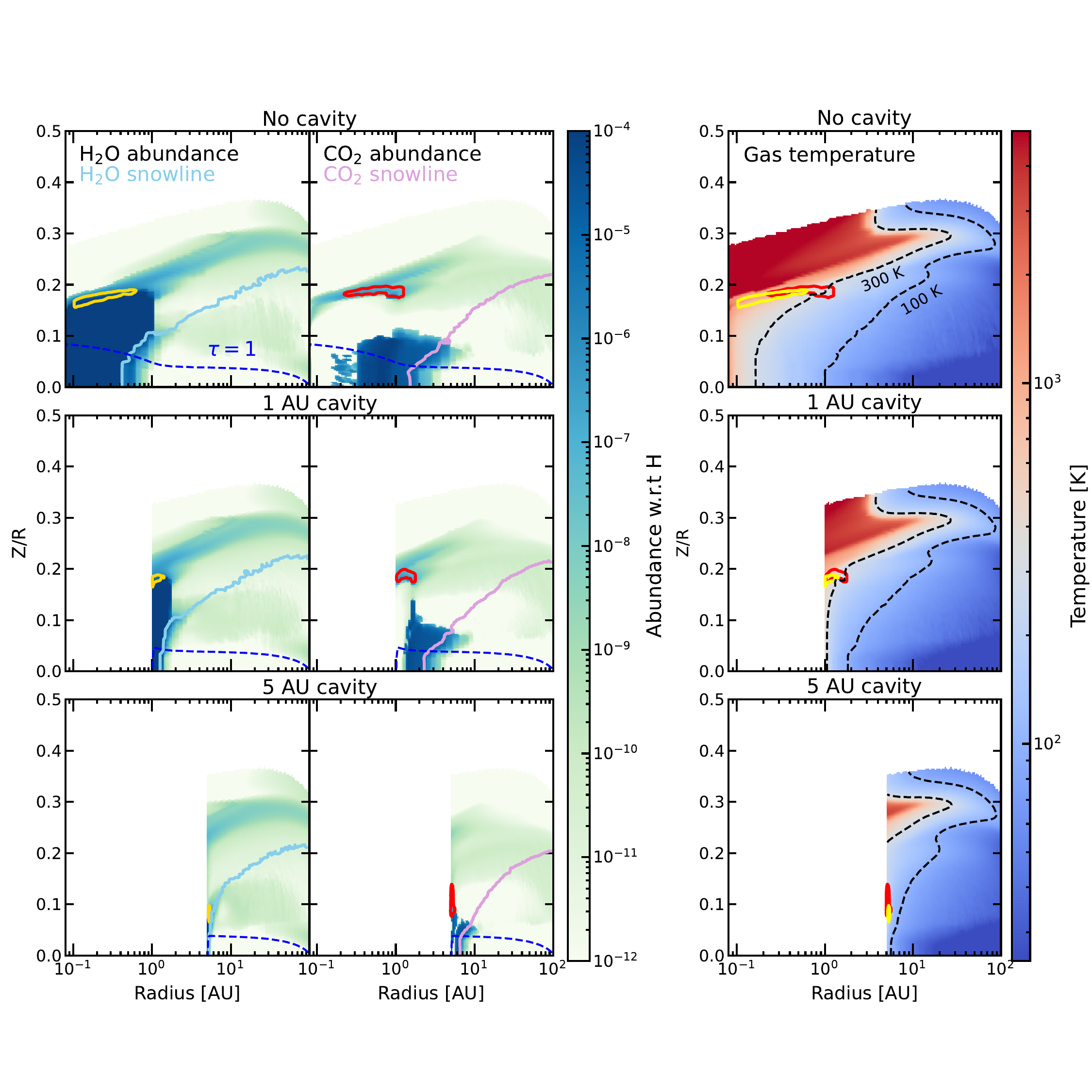}}
    \caption{Abundance maps of \ce{H2O} (left column) and \ce{CO2} (middle column) for the models with no cavity (top row), a 1 AU cavity (middle row) and a 5 AU cavity (bottom row). The \ce{H2O} and \ce{CO2} snowlines (defined as $n_{\rm gas}/n_{\rm ice} = 1$) are indicated with solid light blue and pink lines respectively. The dashed, dark blue line shows the dust $\tau=1$ surface at 15 $\mu$m. The right column shows the gas temperature, with dashed black lines indicating 100 and 300 K. The yellow contours indicate the region in which 90\% of the \ce{H2O} $11_{3,9} - 10_{0,10}$ line emission at 17.22 $\mu$m ($E_{\rm up} = 2438$ K) originates. The red contours indicate the origin of 90\% of the \ce{CO2} $\nu_2 = 1-0$ Q(20) line emission ($E_{\rm up} = 1196$ K).}
    \label{fig:h2o_co2_tgas_grid}
\end{figure*}
\newline
Aside from our fiducial grid of models, we created four additional grids to test the influence of cooling through \ce{H2O} ro-vibrational lines and the luminosity of the central star on the \ce{CO2 /H2O} flux ratio of our model spectra. As such, one additional grid including \ce{H2O} ro-vibrational cooling (which is not included in our fiducial grid, to be consistent with the work by \citealt{bosman2022a, bosman2022b}), is presented in Appendix \ref{sec:app_rovib}. A further three grids are presented in Appendix \ref{sec:app_lum} which contain central sources with luminosities of $L_* = 0.2$, 0.4, and 0.8 $L_\odot$ respectively. We touch on the main results from these tests in Sects. \ref{subsec:rovib} and \ref{subsec:low_lum}, and further details on the specific model setup for these grids can be found in the appendices.  

\begin{figure*}[t]
    \makebox[\textwidth][c]{\includegraphics[scale=0.4]{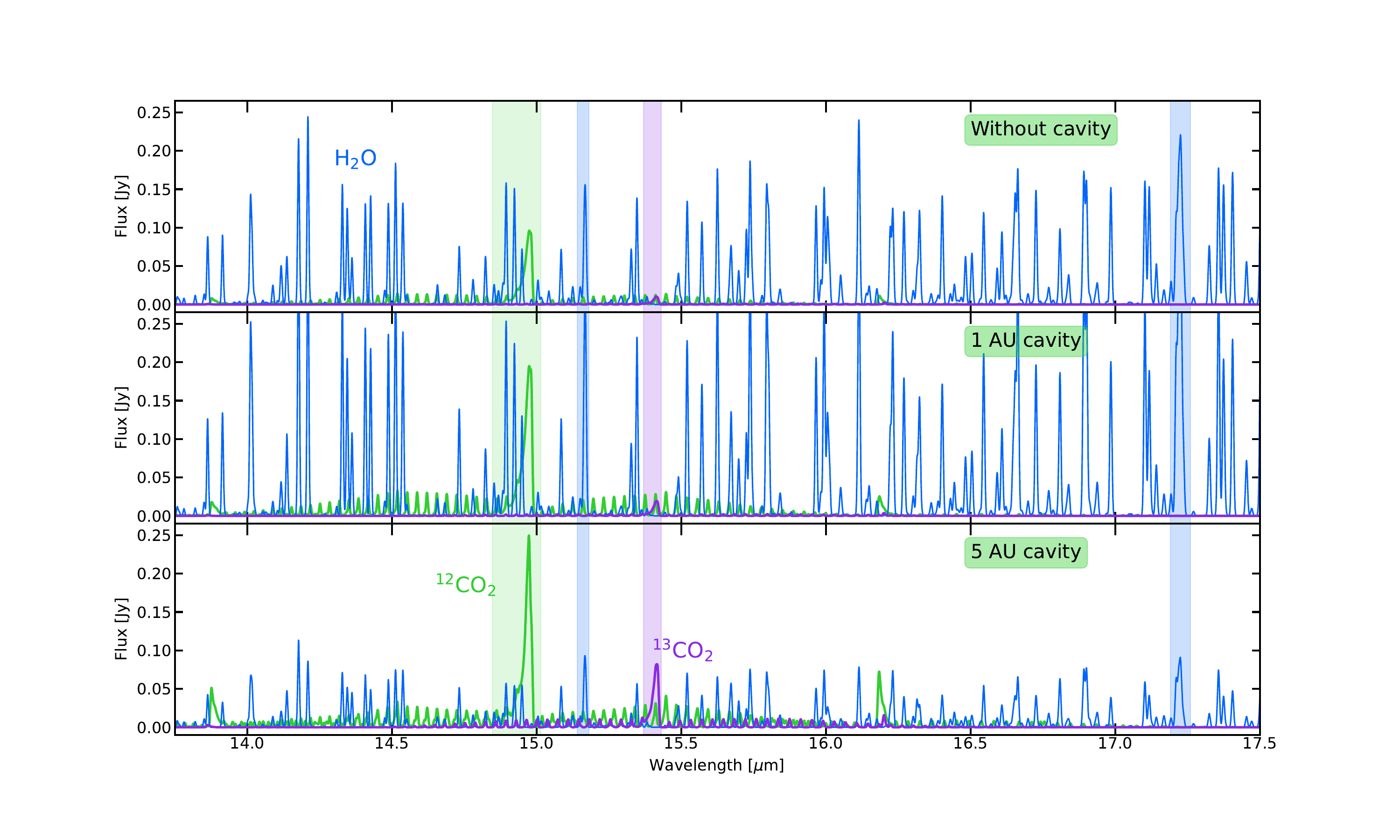}}
    \caption{Generated \ce{H2O}, \ce{^12CO2} and \ce{^13CO2} spectra for the models with no cavity (top row), a 1 AU cavity (middle row) and a 5 AU cavity (bottom row). The vertical colored bars indicate the integration ranges for the \ce{H2O} 15.17 and 17.22 $\mu$m flux (blue), as well as the \ce{^12CO2} and \ce{^13CO2} Q-branches (green and purple respectively).}
    \label{fig:spectra_cav}
\end{figure*}

\begin{figure*}[ht]
    \makebox[\textwidth][c]{\includegraphics[scale=0.4]{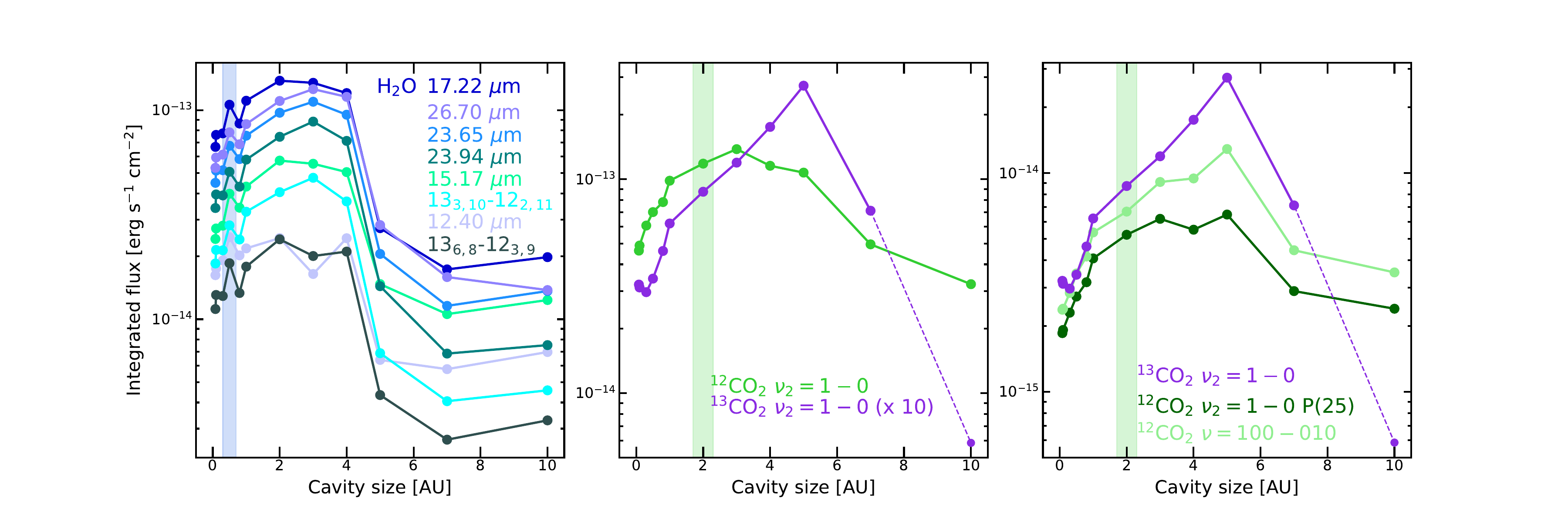}}
    \caption{Integrated fluxes of several \ce{H2O} (left panel),  \ce{^12CO2} and \ce{^13CO2} (middle and right panels) line complexes plotted as a function of cavity size. Properties of these line complexes can be found in Table \ref{tab:Eup}. Nominal \ce{H2O} and \ce{CO2} midplane snowlines in the full disk model are indicated with colored bars. Data points are connected by dashed lines if the line flux drops below $10^{-15}$ erg s$^{-1}$ cm$^{-2}$, indicating that the lines will become more difficult to observe. }
    \label{fig:abs_flux}
\end{figure*}

\section{Results} \label{sec:results} 
\subsection{Temperature and abundance structure} \label{subsec:abu_tgas}

In Fig. \ref{fig:h2o_co2_tgas_grid}, the abundance maps of \ce{H2O} and \ce{CO2} are presented (left and middle columns) for the models with no cavity, a 1 AU and a 5 AU cavity (top, middle and bottom rows respectively). The right hand column presents the corresponding gas temperature. The \ce{H2O} gas abundance maps clearly show how \ce{H2O} is present in three main reservoirs \citep[see also][]{woitke2009}. The inner disk midplane contains the most abundant reservoir of \ce{H2O}. Here, inside the \ce{H2O} snowline, the gas temperature is high enough so that the molecule is effectively formed through gas-phase reactions between OH and \ce{H2}, and the column density is high enough for the molecule to be fully self-shielded from UV radiation. In the warm, upper layers of the disk between $Z/R \sim 0.2-0.3$, the gas temperature is still high enough for gas-phase formation of \ce{H2O} to occur, but due to the lower density, this is now balanced by photodissociation, leading to a slightly lower \ce{H2O} abundance. Finally, a cold photodesorption layer is present in the outer disk around $Z/R \sim 0.1$ where the temperature is too low to effectively form \ce{H2O} via gas-phase reactions or for \ce{H2O} to be thermally sublimated. As such, this reservoir is solely supplied by photodesorped \ce{H2O}, making it the least abundant reservoir. The first two reservoirs are also clearly present for \ce{CO2}, though the midplane reservoir is located further into the disk since \ce{CO2} forms more effectively at lower temperatures and its sublimation temperature is lower. The gap seen between the midplane reservoir and the warm, upper layer of \ce{CO2} is caused by \ce{H2O} UV-shielding, since \ce{CO2} in this region is preferentially destroyed by photodissociation due to its inability to self-shield, and the relatively high gas temperature allows for \ce{H2O} to be preferentially formed \citep[see also][]{bosman2022b}. \\
\newline
The \ce{H2O} and \ce{CO2} snowlines are indicated in the figure with solid light blue and pink lines respectively, and it shows that their nominal midplane positions are at approximately 0.5 and 2 AU respectively for the full disk model. The figure also shows that, as a cavity is introduced into the model, the snowlines will move outwards slightly from their nominal positions. As such, the warm \ce{H2O} and \ce{CO2} midplane reservoirs do not vanish immediately once the cavity size has surpassed their respective nominal snowline positions. Due to this fact, the concept of the `nominal snowline position' is perhaps not the most useful reference to compare the cavity sizes to, since the actual snowline positions in the cavity models will differ. Still, since it is a constant reference point compared to the moving `real' snowlines, we will use it for our comparisons in the rest of this work. \\
\newline
The origin of 90\% of the \ce{H2O} $11_{3,9} - 10_{0,10}$ line ($E_{\rm up}$ = 2438 K) emission at 17.22 $\mu$m and the \ce{CO2} $\nu_2 = 1-0$ Q(20) line ($E_{\rm up}$ = 1196 K) emission at 14.97 $\mu$m are indicated in yellow and red contours respectively. This shows how most of the emission from these molecules is not tracing a deep column of gas, but rather the thin, warm surface layer. Below this warm surface layer, the line emission becomes optically thick, so the IR emission does not probe the deeper reservoirs. The dust $\tau=1$ surface at 15 $\mu$m is indicated with a dashed, dark blue line, showing that the dust becomes optically thick around $Z/R = 0.1$. This is rather low down in the disk, which can be attributed to the fact that the mass fraction of large dust grains is rather large ($f_\ell = 0.999$). These are well settled to the midplane, and thus there is little small dust in the disk upper layers. 

\begin{table*}[ht]
    \centering
    \caption{Line properties}
    \begin{tabular}{c c c c c c}
        \hline \hline
         Molecule & Line complex & Integration range & Transition(s) & $E_{\rm up}$ & $A_{\rm ul}$ \\
          & & [$\mu$m] & & [K] & [s$^{-1}$] \\
         \hline 
         \ce{^12CO2}    & $\nu_2 = 1-0$ Q-branch    & $14.847 - 15.014$\tablefootmark{a} & Q(20) & 1196  & 1.54\\ 
                        & $\nu_2 = 1-0$ P-branch    & $15.435 - 15.460$ & P(25) &  1325  & 0.65\\ 
                        & $\nu = 100 - 010$         & $13.868 - 13.91$  &       &  --  & --\\
        &&&&\\
        \ce{^13CO2}     & $\nu_2 = 1-0$ Q-branch    & $15.37 - 15.43$   &       &  --    & -- \\
        &&&&\\
        \ce{H2O}        & 12.40 $\mu$m & $12.38 - 12.42$  & $17_{4,13} - 16_{3,14}$\tablefootmark{c} &  5781 & 7.7\\
                        &              &                  & $16_{3,13} - 15_{2,14}$\tablefootmark{c} &  4945 & 4.2\\
                        & 15.17 $\mu$m & $15.14 - 15.18$\tablefootmark{b}  & $8_{7,2} - 7_{4,3}$\tablefootmark{d}   &  2288 & 0.06\\
                        &              &                  & $10_{6,4} - 9_{3,7}$                   &  2698 & 0.43\\
                        & 17.22 $\mu$m & $17.19 - 17.26$\tablefootmark{b} & $11_{3,9} - 10_{0,10}$\tablefootmark{c}&  2438 & 0.99\\
                        &              &                                  & $12_{3,9} - 11_{2,10}$                 &  3030 & 2.65\\
                        &              &                                  & $9_{6,4} - 8_{3,5}$                    &  2347 & 0.34\\
                        & 23.65 $\mu$m & $23.615 - 23.66$ & $10_{7,3} - 9_{6,4}$                   &  2956 & 24.4\\
                        &              &                  & $10_{7,4} - 9_{6,3}$                   &  2956 & 24.4\\
                        & 23.94 $\mu$m & $23.91 - 23.96$  & $11_{5,6} - 10_{4,7}$\tablefootmark{d} &  2876 & 10\\
                        &              &                  & $11_{6,6} - 10_{5,5}$                  &  3083 & 16.7\\
                        & 26.70 $\mu$m & $26.67 - 26.74$  & $8_{7,2} - 7_{6,1}$\tablefootmark{d}   &  2288 & 21.48\\
                        &              &                  & $8_{7,1} - 7_{6,2}$                    &  2288 & 21\\
                        & -- & $15.55 - 15.59$ & $13_{3,10} - 12_{2,11}$  & 3474 & 3.04\\
                        & -- & $15.61 - 15.64$ & $13_{6,8} - 12_{3,9}$    & 3953 & 4.03\\
         \hline
    \end{tabular}
    \tablefoot{
    
    For all \ce{H2O}, \ce{^12CO2} and \ce{^13CO2} integrated line fluxes depicted in Figs. \ref{fig:abs_flux}, \ref{fig:rel_flux}, \ref{fig:abs_flux_rv}, \ref{fig:rel_flux_rv}, \ref{fig:abs_flux_lowL} and \ref{fig:rel_flux_lowL} we show here the name/wavelength of the line complex, the wavelength range of over which the flux of the line complex is integrated, transitions within the line complex, and the upper-level energy $E_{\rm up}$ and Einstein-A coefficient $A_{\rm ul}$ of the individual transitions.
    \tablefoottext{a}{\citet{salyk2011}} \tablefoottext{b}{\citet{pontoppidan2010}} \tablefoottext{c}{\citet{banzatti2017}} \tablefoottext{d}{\citet{gasman2023}.}
    }
    \label{tab:Eup}
\end{table*}

\subsection{\texorpdfstring{\ce{H2O} and \ce{^12CO2}}{H2O and 12CO2} spectra} \label{subsec:spectra}

\begin{figure}[t]
    \includegraphics[scale=0.45]{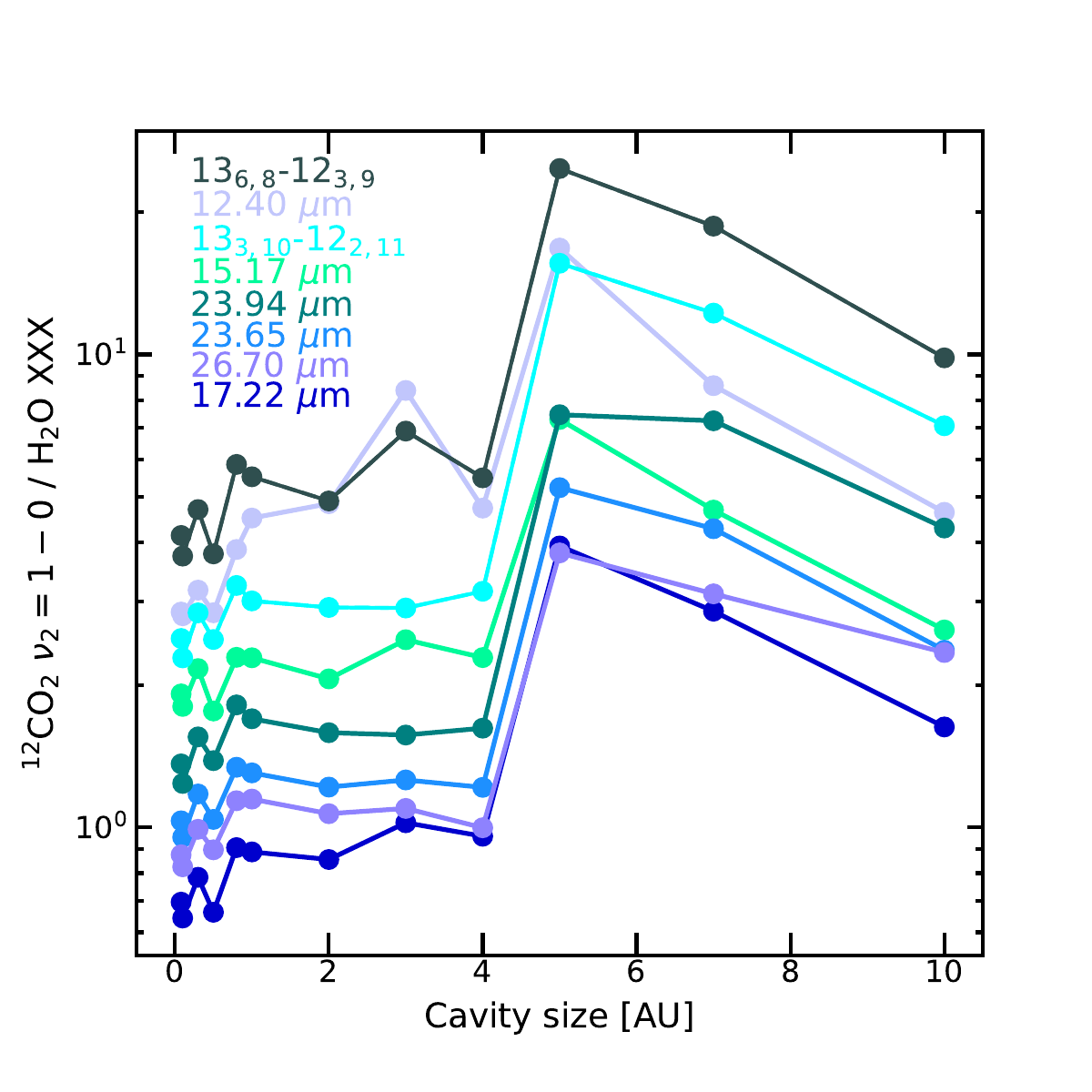}
    \caption{Ratio of the \ce{^12CO2} Q-branch flux and the fluxes of several \ce{H2O} line complexes, plotted as a function of cavity size. Properties of these line complexes can be found in Table \ref{tab:Eup}.}
    \label{fig:rel_flux}
\end{figure}

The generated \ce{H2O}, \ce{^12CO2} and \ce{^13CO2} continuum-subtracted spectra are presented in Fig. \ref{fig:spectra_cav} for the models with no cavity, a 1 AU cavity and a 5 AU cavity, which are shown in the top, middle and bottom panels respectively. This figure shows that the spectrum for the full disk is dominated by \ce{H2O} emission, and becomes \ce{CO2}-dominated only once a large-enough cavity is introduced. This thus confirms the plausibility of our initial hypothesis, stating that the observation of bright \ce{CO2} emission in the mid-IR spectrum of a disk could be caused by that disk containing a small, inner cavity.\\
\newline
To summarize the evolution of the spectra across all cavity sizes concisely, the integrated fluxes of several \ce{H2O} line complexes as well as the \ce{^12CO2} and \ce{^13CO2} Q-branches are plotted as a function of cavity size in Fig. \ref{fig:abs_flux}. 
Properties of the line complexes whose fluxes are shown in these figures, such as the wavelength range over which the flux is calculated and the upper-level energy $E_{\rm up}$ of the most prominent transitions within that wavelength range, can be found in Table \ref{tab:Eup}. For \ce{^12CO2} and \ce{^13CO2} we chose to calculate the fluxes of their respective $\nu_2 = 1-0$ Q-branches, since they are these molecules' most prominent features in the mid-IR wavelength range. For \ce{^12CO2}, the integration range used is the same as in \citet{salyk2011} and \citet{anderson2021}. We also calculate the fluxes of the \ce{^12CO2} $\nu = 100-010$ 13.9 $\mu$m hot band and the $\nu_2 = 1-0$ P(25) line, since these lines are more comparable in flux to the \ce{^13CO2} Q-branch. For \ce{H2O}, we calculate the fluxes of the 15.17 and 17.22 $\mu$m line complexes as was also done for \textit{Spitzer} data in \citet{pontoppidan2010}. We also include a blend of two lines with higher $E_{\rm up}$ values at 12.4 $\mu$m that are also analysed in \citet{banzatti2017}. Additionally, we include the line complexes at 23.94 and 26.70 $\mu$m as they are also discussed in the analysis of JWST-MIRI data of \ce{H2O}-rich source Sz 98 by \citet{gasman2023}, and they have slightly higher $A_{\rm ul}$ values. Finally, we also include a line complex at 23.65 $\mu$m that is likely to be detected in \ce{H2O}-rich sources with JWST-MIRI (see Sect. \ref{subsec:obs_cav}) and two individual, unblended lines at $\sim$ 15.6 $\mu$m. All of the lines considered here are (marginally) optically thick in all models and thus their fluxes scale with the temperature and emitting area, rather than scaling linearly with the total number of molecules $\mathcal{N}_{\rm tot} = N \pi R^2$. Analyzing the evolution of these fluxes as a function of cavity size gives some insight into the evolution of the spectrum as a whole as a cavity is introduced in a disk. \\
\newline
All depicted lines can be seen to increase in flux as a small cavity ($\lesssim$ 3-4 AU) is introduced (Fig. \ref{fig:abs_flux}). This seems slightly counter-intuitive, since there is now less molecular gas present in the inner region of the disk, but this same effect has also been seen in other works \citep[see, e.g., analysis by][]{antonellini2016, woitke2016, woitke2018}. The explanation for this most likely lies with the fact that the cavity wall is now directly irradiated by the star, and thus also directly heated. As discussed above, the snowlines are pushed outwards as a result of this. Additionally, since the cavity wall is now directly exposed, more of the IR emission will be tracing the cavity wall directly, instead of the disk surface. Since the cavity wall area grows as $R_{\rm cav}^2$, the emission is now tracing a larger surface area of warm gas, causing the line flux to increase. \\
\newline
However, for larger cavity sizes ($\gtrsim$ 5 AU), the line flux drops off steeply. At these cavity sizes, the direct irradiation of the cavity wall can no longer heat the gas enough to produce a large column of gas. Only a very small region right at the cavity wall remains hot enough for gas-phase formation of \ce{CO2} and \ce{H2O} and as such, the line flux drops off. This can also be quite clearly seen when considering the emitting regions shown in Fig. \ref{fig:h2o_co2_tgas_grid}. For smaller cavity sizes, the molecular emission traces a warm, thin surface layer of gas, but for a cavity size of 5 AU and larger this layer is absent. This observed trend in flux (the initial increase followed by a sharp drop-off) is consistent with previous work by \citet{antonellini2016, woitke2018}. Interestingly, the work by \citet{anderson2021} shows a different behavior. In their work, they vary the inner radius of their model between 0.2 and 1 AU and observe a substantial decrease in molecular line flux, instead of the initial increase in flux that would be expected for such small cavity sizes. \\
\newline
Comparing the evolution of the \ce{CO2} and \ce{H2O} line fluxes in Fig. \ref{fig:abs_flux}, we can see that the \ce{H2O} line fluxes drop off at smaller cavity sizes than the \ce{CO2} fluxes ($\sim$4 AU vs. $\sim$5-6 AU). This is to be expected, since \ce{H2O} has a higher binding energy, and is formed more efficiently at higher temperatures than \ce{CO2}, thus it is located further into the disk. As such, we can see from Fig. \ref{fig:spectra_cav} that this can lead to a spectrum that was \ce{H2O}-dominated when it had no cavity to a spectrum becoming \ce{CO2}-dominated once a large-enough cavity is introduced. This is in line with what was expected from the scenario that was proposed. However, we expected this switch from \ce{H2O}-dominated to \ce{CO2}-dominated to occur when the disk had a cavity that extended beyond the nominal \ce{H2O} snowline position, but not beyond the nominal \ce{CO2} snowline position. From our modelling work, this clearly does not hold up. This is best seen in Fig. \ref{fig:rel_flux}
, where the flux of the \ce{^12CO2} Q-branch is now divided by the \ce{H2O} line fluxes from the left panel of Fig. \ref{fig:abs_flux}. This figure clearly shows that it takes a cavity of 5 AU or larger for the relative \ce{CO2 /H2O} flux to reach its maximum, i.e. for the spectrum to become \ce{CO2}-dominated. Still, for cavity sizes of 5 AU and smaller, the relative \ce{CO2 /H2O} fluxes do show an overall upwards trend
, which shows that this scenario produces the expected outcome, namely that it becomes more likely to clearly distinguish the \ce{CO2} Q-branch from the \ce{H2O} lines in this wavelength range as a small cavity is introduced into the disk. 
It should be noted that Fig. \ref{fig:rel_flux} shows a noticeable downturn in the relative \ce{CO2 /H2O} flux for cavity sizes larger than 5 AU. This is caused by both the \ce{H2O} and \ce{CO2} line fluxes rapidly dropping off for these models, and thus the relative fluxes begin to equalize.\\ 

\subsection{\texorpdfstring{\ce{^13CO2}}{13CO2} spectrum} \label{subsec:13co2}
Aside from \ce{^12CO2}, the spectrum of \ce{^13CO2} and the integrated flux of its Q-branch are also presented in Fig. \ref{fig:spectra_cav} and in the middle and right panels of Fig. \ref{fig:abs_flux}, respectively. From these two figures, it is clear that the \ce{^13CO2} Q-branch has a similar intensity to the surrounding \ce{^12CO2} P-branch lines for cavity sizes up to $\sim 2$ AU \citep[see also][]{bosman2017, bosman2022b}. For cavity sizes of 4-5 AU, the flux of the \ce{^13CO2} Q-branch reaches its peak and becomes roughly 2 times brighter than the \ce{^12CO2} P-branch lines. For even larger cavity sizes, the flux drops off steeply again. The peak in flux of the \ce{^13CO2} Q-branch at cavity sizes of 4-5 AU could indicate that the emission in starting to trace deeper into the disk. \\
\newline
When looking at the \ce{CO2} emitting region in the bottom panels of Fig. \ref{fig:h2o_co2_tgas_grid}, this indeed seems to be the case. This is further supported by the flux of the \ce{^12CO2} $\nu = 100 - 010$ hot band that is shown in the right panel of Fig. \ref{fig:abs_flux}. \citet{bosman2022b} show that, if the $\nu = 100 - 010$ hot band is brighter than the $\nu_2 = 1-0$ P(19)-P(27) lines, this is indicative of the emission tracing deep, warm \ce{CO2}. In our models, we see that this is most prominently the case around cavity sizes of 4-5 AU. This analysis also shows that the \ce{^13CO2} emission will be best observable right after the point of transition from a \ce{H2O}-dominated to a \ce{CO2}-dominated spectrum, and that it has the potential to become quite prominent compared to the \ce{^12CO2} P-branch lines. \ce{^13CO2} has been detected in the gas-phase for the first time in the GW Lup disk by \citet{grant2023}. In the context of the hypothesis that this disk has a small cavity, this detection is in line with our expectations. The authors compare the peak intensity of the \ce{^13CO2} Q-branch to that of the \ce{^12CO2} P(27) line, finding a ratio of 1.4. In our work, we find the ratio of peak intensities for models with cavity sizes of 4-5 AU to be similar to the finding by \citet{grant2023}. 

\subsection{Inclusion of \texorpdfstring{\ce{H2O}}{H2O} ro-vibrational cooling} \label{subsec:rovib}
\begin{figure}[b]
    \includegraphics[width=\hsize]{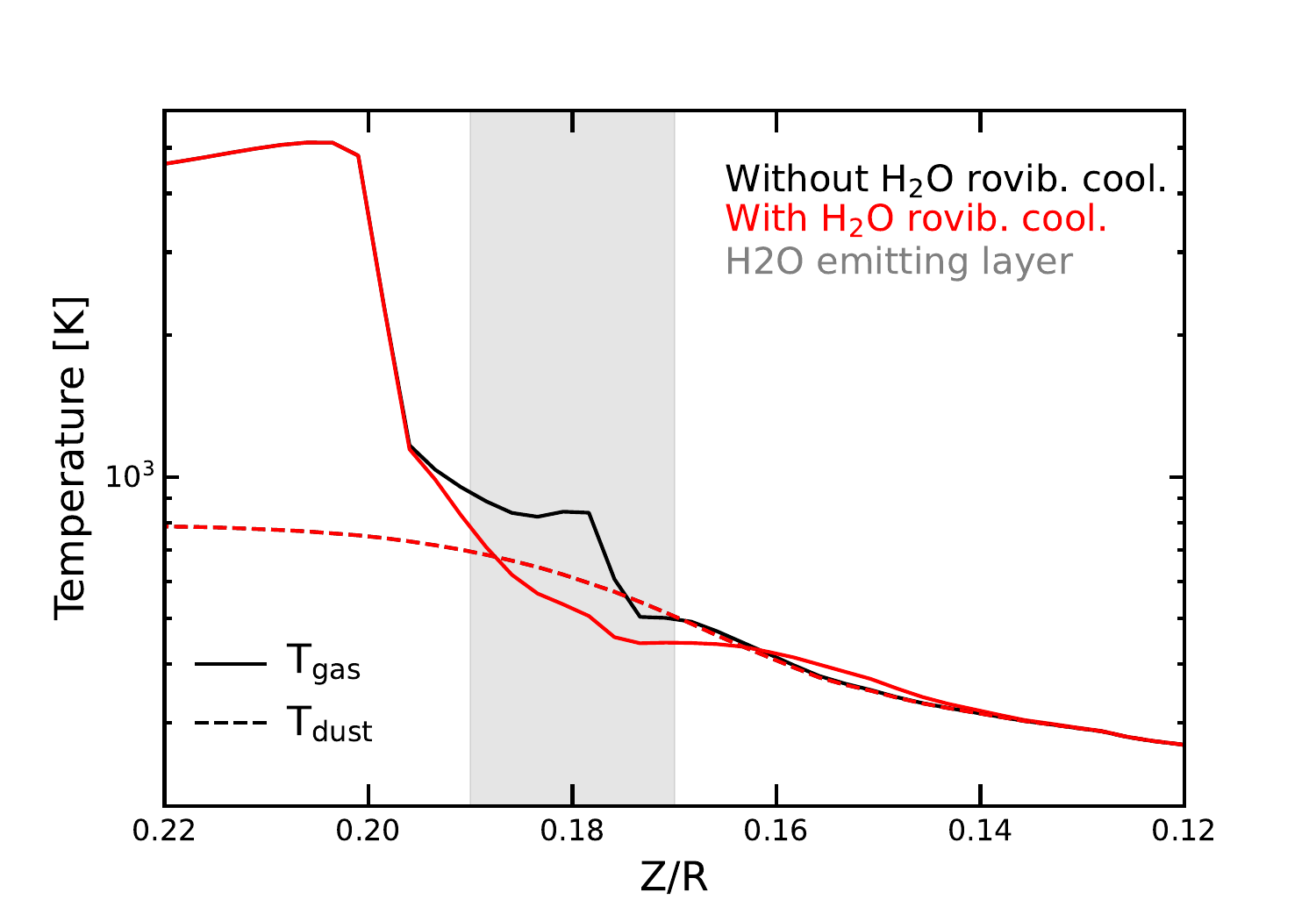}
    \caption{Gas and dust temperatures (solid and dashed lines respectively) as a function of height in the disk $Z/R$ for the models with and without \ce{H2O} ro-vibrational cooling, indicated in black and red respectively. Both are full disk models without a cavity. The height in the disk at which \ce{H2O} emits is indicated in grey. The vertical cut through the disk is taken at a distance of 0.3 AU.}
    \label{fig:tgas_cut03}
\end{figure}

When computing the spectra, we realised that cooling by mid-IR and near-IR molecular lines can significantly impact the \ce{H2O} fluxes. Specifically, cooling by the ro-vibrational transitions of \ce{H2O} has not been included in the results shown so far. This was done to be consistent with work by \citet{bosman2022a, bosman2022b}, who do not include \ce{H2O} ro-vibrational cooling since this yielded low fluxes compared with observations (A. Bosman, private communication). Indeed, we find that the inclusion of this cooling lowers the \ce{H2O} fluxes by a factor of 2-3, since the disk gas temperature is lowered precisely in the layer from which \ce{H2O} predominantly emits in the mid-IR. Since this effect is highly localised to a specific layer in the inner disk, \ce{H2O} fluxes at longer wavelengths, such as those in the far-IR or ALMA wavelength range, will likely be much less affected, if at all. 
The effect on the temperature structure can be seen in Fig. \ref{fig:tgas_cut03}, where the gas and dust temperatures (solid and dashed lines respectively) are shown for a vertical cut of the disk taken at 0.3 AU. The model without \ce{H2O} ro-vibrational cooling (indicated in black) is hotter by, at most, a factor of 2 compared to the model in which this cooling is included (indicated in red). This increase in temperature is limited to a narrow range of disk heights that coincides with the \ce{H2O} emitting layer (indicated in grey). The inclusion of \ce{H2O} ro-vibrational transitions in the heating and cooling balance allows the gas more ways to cool in this layer, hence the lower temperature. \\
\newline
The fact that this particular layer is sensitive to such effects is not surprising. Looking at the gas temperature maps for our models in the right column in Fig. \ref{fig:h2o_co2_tgas_grid}, we can see that the \ce{H2O} emitting region lies just below the hot, $T$ > 1000 K disk atmosphere where most gas is atomic. In the layer below this hot atmosphere, which our molecular emission is tracing, the gas is molecular, but the gas and dust are still thermally decoupled, as can also be seen in Fig. \ref{fig:tgas_cut03}. As such, the gas temperature in this layer is particularly sensitive to changes in the thermochemistry. This is also seen in \citet{bosman2022a}, where the authors show how the inclusion of \ce{H2O} UV-shielding and chemical heating leads to changes in gas temperature in this layer. \\
\newline
To investigate the effects of \ce{H2O} ro-vibrational cooling, we created an additional grid in which we did include \ce{H2O} ro-vibrational cooling. The analysis shown in the previous three subsections was repeated for this grid and is presented in Appendix \ref{sec:app_rovib}. Comparisons between the \ce{H2O} and \ce{CO2} abundance maps, the gas temperature map and the generated spectra from the two grids are shown in Figs. \ref{fig:h2o_co2_rv_nonrv}, \ref{fig:tgas_rv_nonrv}, and \ref{fig:spectra_subset}. The integrated fluxes of several \ce{H2O}, \ce{^12CO2} and \ce{^13CO2} lines (analogous to Fig. \ref{fig:abs_flux}) are presented in Fig. \ref{fig:abs_flux_rv}, and the relative \ce{CO2 /H2O} fluxes of these lines (analogous to Fig. \ref{fig:rel_flux}) are presented in Fig. \ref{fig:rel_flux_rv}.\\
\newline
The resulting decrease in \ce{H2O} flux from the inclusion of \ce{H2O} ro-vibrational cooling is of the order of a factor 2-3. This may not be a very significant change by itself, but when the \ce{H2O} flux is compared to the \ce{CO2} flux, the spectra change from being dominated by the \ce{H2O} lines to being dominated by the \ce{CO2} Q-branch feature. Thus, whereas our fiducial models (the models without \ce{H2O} ro-vibrational cooling) were \ce{H2O}-dominated until a large-enough cavity was introduced, the models with \ce{H2O} ro-vibrational cooling are always \ce{CO2}-dominated. Regardless, when a cavity is introduced, both grids show a similar trend. As can be seen in Figs. \ref{fig:spectra_subset} and \ref{fig:abs_flux_rv}, the fluxes first slightly increase when a small cavity is introduced, which is then followed by a decrease in flux once that cavity becomes large enough. This shows that the trends in flux found in this work are robust, but the model assumptions for the full disk spectrum, determining whether the model is \ce{H2O}- or \ce{CO2}-dominated, are important.

\subsection{Lower-luminosity central sources} \label{subsec:low_lum}
\begin{figure}[t]
    \includegraphics[scale=0.45]{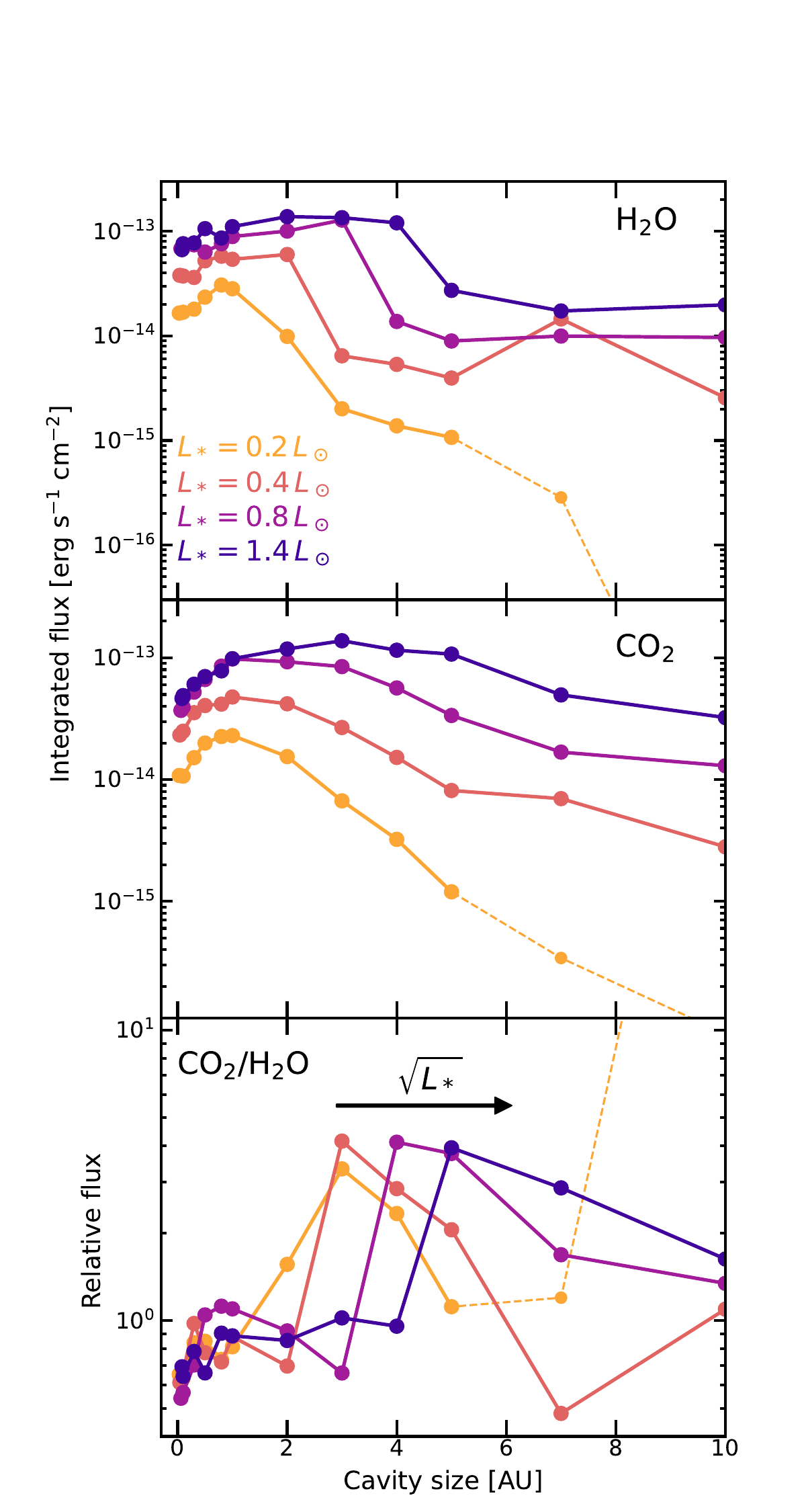}
    \caption{Flux of the 17.22 $\mu$m \ce{H2O} line complex (top), the \ce{CO2} Q-branch (middle) and their ratio (bottom) for the grids with $L_* = 0.2, 0.4, 0.8$ and $1.4\,L_\odot$.}
    \label{fig:flux_lowL}
\end{figure}

To better compare our models to the \ce{CO2}-only source GW Lup, we also present the results obtained from models that have a lower-luminosity input spectrum than our fiducial AS 209 spectrum in Appendix \ref{sec:app_lum}. \ce{H2O} ro-vibrational cooling was not included, just like our fiducial grid. The input spectra used for these grids have $L_* = 0.2, 0.4$ and $0.8\,L_\odot$ compared to the $L_* = 1.4\,L_\odot$ of the fiducial grid, to ensure that we cover a decent luminosity range. For reference, GW Lup has a luminosity of roughly 0.3 $L_\odot$ \citep{alcala2017, andrews2018}. We started our analysis with our fiducial, AS 209 model rather than a lower-luminosity model more comparable to GW Lup in order for our fiducial models to be directly comparable to the work by \citet{bosman2022a, bosman2022b}. A comparison of the \ce{H2O} and \ce{CO2} abundance maps as well as the gas temperature map between all four luminosity grids is presented in Figs. \ref{fig:h2o_comp_lowL}, \ref{fig:co2_comp_lowL}, and \ref{fig:tgas_comp_lowL}. The generated \ce{H2O}, \ce{^12CO2} and \ce{^13CO2} spectra are presented in Fig. \ref{fig:spectra_subset_lowL}, the integrated fluxes of several \ce{H2O}, \ce{^12CO2} and \ce{^13CO2} lines (analogous to Fig. \ref{fig:abs_flux}) are presented in Fig. \ref{fig:abs_flux_lowL}, and the relative \ce{CO2 /H2O} fluxes of these lines (analogous to Fig. \ref{fig:rel_flux}) are presented in Fig. \ref{fig:rel_flux_lowL}. \\
\newline
The spectra of these models are all, just like the fiducial model, \ce{H2O}-dominated when no cavity is present, and they all become \ce{CO2}-dominated once a large enough cavity is introduced. So, the lower-luminosity input spectrum and resulting decrease in global disk gas temperature by itself is not enough to make these models \ce{CO2}-dominated. The overall trend they follow is also the same as what was described above: the \ce{H2O} and \ce{CO2} line emission initially becomes brighter as a small cavity is introduced, and once that cavity becomes significantly larger than the molecule's nominal snowline position, the emission drops off sharply. This is shown using the 17.22 $\mu$m \ce{H2O} flux, the \ce{CO2} Q-branch flux and the ratio of these two for the grids of all four luminosities in Fig. \ref{fig:flux_lowL}. This figure clearly shows that the drop in line flux from \ce{H2O} and \ce{CO2} occurs at smaller cavity sizes for models with a lower-luminosity input spectrum. The \ce{CO2 /H2O} flux ratio shows a similar trend between all four models: there is an upwards trend for smaller cavity sizes and a downturn for larger cavity sizes. However, as one would expect, the lower-luminosity models reach their peak \ce{CO2 /H2O} flux ratio (and thus become \ce{CO2}-dominated) at smaller cavity sizes (e.g. $\sim$ 2-3 AU for a $L_* = 0.2 \,L_\odot$ source vs. $\sim$ 4-5 AU for a $L_* = 1.4 \,L_\odot$ source). Still, these values lie well beyond each model's nominal midplane \ce{CO2} snowline position. We discuss the consequences and possible causes of this in Sect. \ref{subsec:cav_snowline}. Additionally, from Fig. \ref{fig:flux_lowL} we can see that the \ce{CO2} and \ce{H2O} line fluxes approximately scale with the stellar luminosity as $\sqrt{L_*}$, as is also found by Tabone et al. (2023, subm.). This is shown more clearly in Fig. \ref{fig:flux_lowL_rescaled}, which shows the same data as Fig. \ref{fig:flux_lowL}, but with the x-axis re-scaled by a factor $\sqrt{L_*}$. In this figure, the peaks of the line fluxes now all roughly coincide, whereas they are shifted with respect to each other in Fig. \ref{fig:flux_lowL}. From all of this, we can conclude that our main findings from the fiducial grid also hold for lower-luminosity sources.


\section{Discussion} \label{sec:discussion}

\subsection{A cavity as an explanation for strong \texorpdfstring{\ce{CO2}}{CO2} emission} \label{subsec:cav_snowline}
In this work, we test the scenario proposed by \citet{grant2023} to explain the strong \ce{CO2} emission that some IR spectra of T Tauri disks show: the presence of a small, inner cavity, located between the \ce{H2O} and \ce{CO2} snowlines, could suppress the \ce{H2O} emission, thus leading the source's spectrum to show a large \ce{CO2 /H2O} flux ratio. With this modeling work, we show that this is indeed a possibility: a spectrum that was \ce{H2O}-dominated when it had no cavity can become \ce{CO2}-dominated instead once a large-enough cavity is introduced. We expected the cavity size needed for a spectrum to switch to be larger than the nominal position of the \ce{H2O} snowline, but not larger than that of the \ce{CO2} snowline. From our modeling, however, we consistently find that this switch occurs for a cavity size that is well beyond the nominal position of the \ce{CO2} snowline in a full disk. Some discussion on possible causes and consequences is thus warranted. \\
\newline
Some possible causes can be understood by looking back at Fig. \ref{fig:h2o_co2_tgas_grid}. As was already pointed out in Sect. \ref{subsec:abu_tgas}, the reservoir of warm molecular gas does not disappear immediately once a cavity exceeding the position of its snowline is introduced. For example, the midplane \ce{H2O} snowline is clearly within 1 AU in the model without a cavity (top left panel of Fig. \ref{fig:h2o_co2_tgas_grid}), but when a 1 AU cavity is introduced, the warm \ce{H2O} reservoir near the cavity wall is still quite prominent. Clearly, the snowlines move outwards when a cavity is present. As such, it makes sense that the emission also remains prominent. Additionally, we see from Fig. \ref{fig:h2o_co2_tgas_grid} that the \ce{H2O} and \ce{CO2} emission originates from a thin, warm surface layer at $Z/R \sim 0.2$ in the disk. Looking at the right column of this figure, we can see that this warm layer remains present in the disk until a cavity size of roughly 5 AU, when we also see that the molecular emission is starting to trace regions slightly deeper down in the disk. The continued presence of this warm layer, combined with the continued presence of molecular gas near the cavity wall, likely explains why the molecular emission remains bright, even though a cavity far larger than the nominal \ce{CO2} snowline position has been introduced.  \\
\newline
To turn an otherwise \ce{H2O}-dominated spectrum into a \ce{CO2}-dominated one, the cavity that needs to be introduced must be larger than was initially expected. This makes it less likely for this phenomenon to be the sole explanation for the existence of the \ce{CO2}-only sources found by {\it Spitzer}. Specifically for sources with higher luminosity, for which cavity sizes of 4-5 AU are needed, the presence of such a cavity could more easily be ruled out observationally (see Sect. \ref{subsec:obs_cav}). Additionally, our modeling work shows that small, local temperature changes can have a big impact on a disk's \ce{CO2 /H2O} flux ratio. The difference in temperature caused by the inclusion of \ce{H2O} ro-vibrational cooling is, at most, only a factor of two, and it is localised to a very specific layer in the disk, yet that is enough to switch from a \ce{H2O}-dominated spectrum to a \ce{CO2}-dominated one. Global temperature changes on the other hand, for example those caused by changes in stellar luminosity, do not seem to have much of an impact on the \ce{CO2 /H2O} flux ratio, instead impacting both molecules' fluxes equally. 

\subsection{Observing a small inner cavity} \label{subsec:obs_cav}

To confirm this scenario producing a large \ce{CO2 /H2O} flux ratio, one would want to observe a small cavity in sources that are known to have a \ce{CO2}-dominated mid-IR spectrum. Here, we discuss some methods that could be used to do so. First, our modeling work has shown that the cavity needed to make a source's spectrum \ce{CO2}-dominated can be quite large, $\sim 5$ AU. With sufficiently high angular resolution mm observations, for example comparable to that of the DSHARP program (\citealt{andrews2018}; $0\farcs{035}$, $\sim 5$ AU diameter for a source at 150 pc), such a cavity's presence could be confirmed or ruled out. A source's SED can also provide a measure of any inner cavity present in the disk, though this is in part degenerate with the exact dust properties \citep[see, e.g.][]{woitke2016}. In this work, we assume an inner cavity to be completely devoid of dust, but this does not need to be the case. If there was a sufficient amount of small grains left in the cavity, this could mask its presence in the SED. \\ 
\newline
Velocity-resolved line profiles, for example those of the fundamental CO ro-vibrational transitions at 4.7 $\mu$m, could provide a better measure of the inner gas radius \citep[e.g.][]{brown2013, banzatti2022}. This could be obtained with the VLT CRIRES+ or the Keck NIRSPEC spectrographs. Such data exist for many disks, but not yet for most \ce{CO2}-rich sources. Finally, one could attempt to resolve the inner dust disk directly through IR interferometric observations with, for example, VLTI GRAVITY at 2 $\mu$m or MATISSE in the $L$, $M$ and $N$ bands \citep[see, e.g.][]{varga2021, bohn2022, gravitycollab2021}.  In this case, however, the distinction between an inner cavity (no inner disk present) and an inner gap (with inner disk present) may not be clear. Additionally, this technique may be limited to the brightest sources due to the limited sensitivity of these instruments.

\begin{figure}[t]
    \includegraphics[scale=0.45]{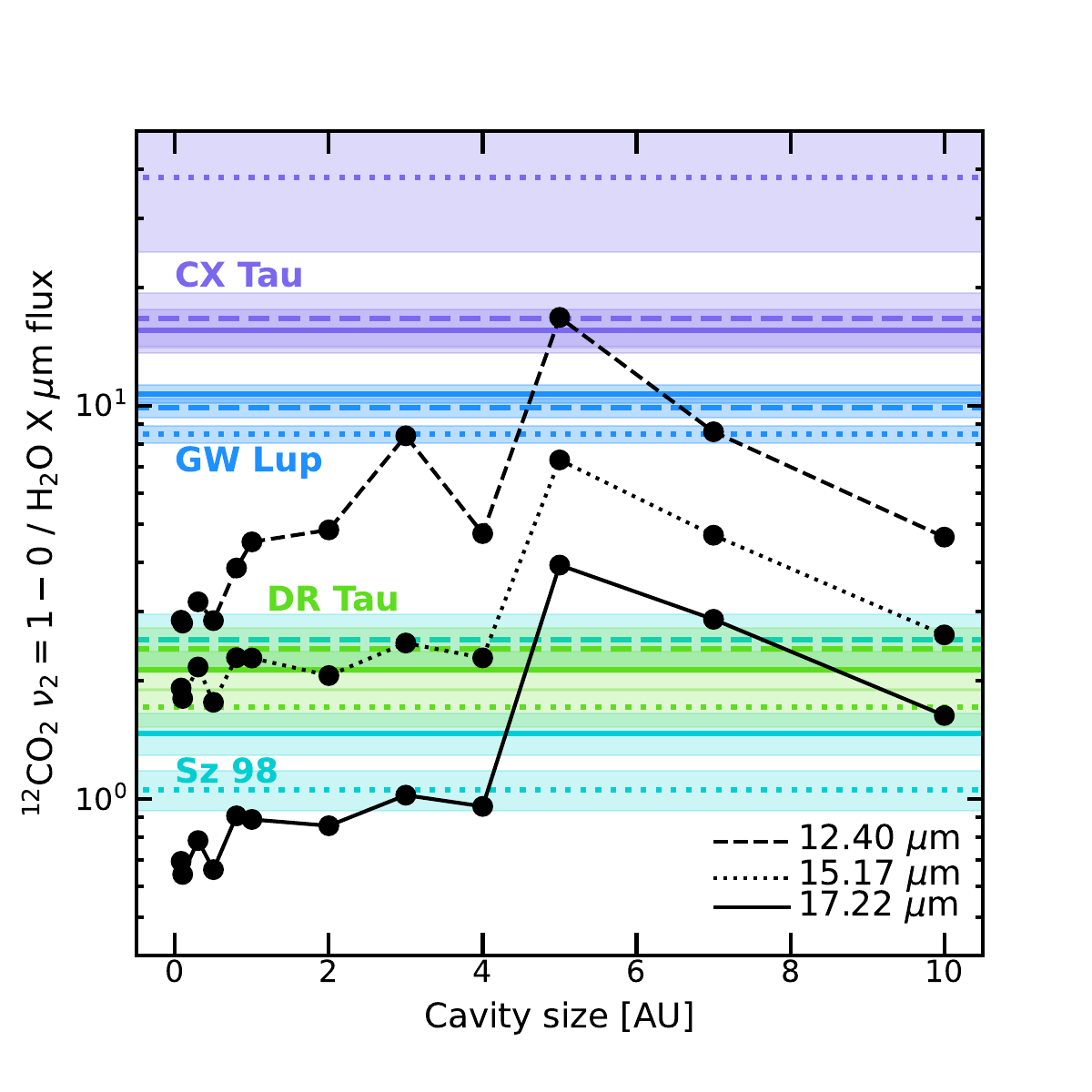}
    \caption{The ratio of the \ce{^12CO2} Q-branch flux and the fluxes of the 12.40, 15.17 and 17.22 $\mu$m \ce{H2O} line complexes from our fiducial grid of models, plotted as a function of cavity size, is shown with dashed, dotted and solid black lines respectively. Vertical colored lines, with bars indicating the uncertainty, indicate the same \ce{CO2 /H2O} flux ratios calculated for JWST-MIRI observations of GW Lup \citep{grant2023}, Sz 98 \citep{gasman2023}, DR Tau (Temmink et al., in prep.) and CX Tau (Vlasblom et al., in prep.), all taken as a part of the MINDS program. }
    \label{fig:rel_flux_obs}
\end{figure}

\subsection{Comparison to observations} \label{subsec:comp_obs}

Having determined the relative \ce{CO2 /H2O} fluxes in our models, we can compare them to some disks recently observed with JWST-MIRI as part of the MINDS program. For four sources, the integrated fluxes of the \ce{H2O} 12.40, 15.17 and 17.22 $\mu$m line complexes, as well as the \ce{^12CO2} Q-branch, are calculated and their ratios are presented as vertical colored lines in Fig. \ref{fig:rel_flux_obs}. Two of these sources are clearly \ce{H2O}-rich: DR Tau (Temmink et al., in prep.; see also \citealt{salyk2008}) and Sz 98 (\citealt{gasman2023}; see also \citealt{vandishoeck2023, kamp2023}). The other two are clearly \ce{CO2}-rich: GW Lup \citep{grant2023} and CX Tau (Vlasblom et al., in prep.). Plotted as black lines are these same flux ratios calculated from our fiducial grid of models, as presented in Fig. \ref{fig:rel_flux}. The \ce{CO2 /H2O} flux ratios of DR Tau and Sz 98 are roughly in line with our \ce{H2O}-dominated models, those that have no or only a small cavity. The flux ratios of GW Lup and CX Tau are even larger than those of our \ce{CO2}-dominated models, indicating that these sources have even fainter \ce{H2O} emission than what our models predict. This could be an indication that an additional explanation beyond just the presence of a cavity may be needed for these high flux ratios (see Sect. \ref{subsec:altexpls}). Still, it is clear that GW Lup and CX Tau are the best candidates for having a small, inner cavity. \\
\newline
GW Lup is an M1.5 star at a distance of 155 pc with a luminosity of $L_*=0.33\,L_\odot$ that has been observed as part of the DSHARP program \citep[see][]{alcala2017, andrews2018}. Its dust disk is rather extended with an outer radius of $\sim$ 120 AU in mm continuum and its radial intensity profile does not show a decrease towards the center, which would hint at the possible presence of a dust cavity \citep{huang2018}. Of all of the sources presented in the DSHARP survey, there are a few sources that do show a decrease in intensity towards the center of their disks, however from their radial intensity profiles we can estimate these possible cavities or gaps to be roughly 10 AU in radius. Since GW Lup has a stellar luminosity roughly consistent with the two lowest-luminosity grids presented in this work, we can estimate that it would need a cavity of approximately 3 AU in size to explain its bright \ce{CO2} emission. Such a cavity would likely not be visible at the distance GW Lup is located, even at DSHARP resolution. Instead, velocity-resolved CO data or VLTI observations could provide further insights.  \\
\newline
CX Tau is an M2.5 star at a distance of 128 pc with a luminosity of $L_*=0.22\,L_\odot$ \citep{herczeg2014}. Its dust disk is very compact, having a 68\% dust radius of only $\sim$ 15 AU. Its gas disk, on the other hand, is much more extended ($\sim$ 75 AU), strongly pointing to radial drift being an important mechanism in this disk  \citep{facchini2019}. CX Tau was observed with {\it Spitzer} \citep[see][]{najita2007, furlan2011} and though its molecular features were never analyzed in detail, it can certainly be classified as a \ce{CO2}-only source. Following our modeling, a cavity of roughly 2 AU in radius could explain its high \ce{CO2} flux. Interestingly, this source has been classified as a transition disk in the past based on its lack of near-IR excess, though this classification does not hold when considering color criteria (compare \citealt{najita2007} and \citealt{furlan2011}). However, high-resolution ALMA data (taken at the same resolution as the DSHARP program) do not show any signs of an inner cavity. Analysis by \citet{facchini2019} shows that, when adding a parameter $R_{\rm trunc}$ to their fit of the disk's intensity profile within which the dust continuum intensity is set to zero, an upper limit of $R_{\rm trunc} = 0.54$ AU encloses 95\% of their MC realizations. As such, it is unlikely that the disk has an inner cavity much larger than this. From our models, we can conclude that such a cavity would not be enough to explain the high \ce{CO2} flux and relatively low \ce{H2O} flux. 

\begin{figure}[b]
    \includegraphics[scale=0.38]{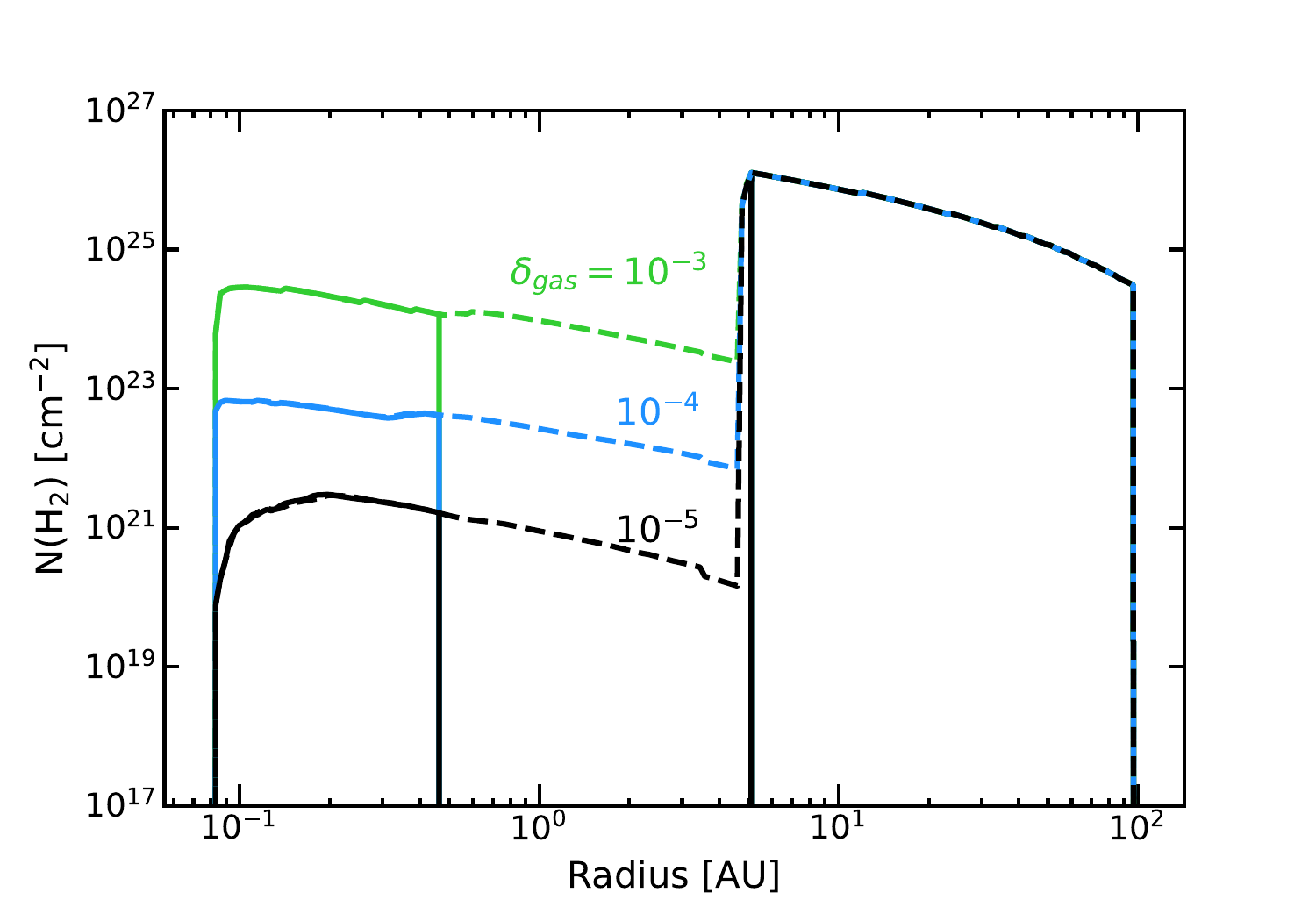}
    \caption{\ce{H2} column density, integrated until $\tau_{\rm dust} = 1$ at 15 $\mu$m, as a function of disk radius for the models with an 0.5 AU inner disk (solid lines) and gas inside a 5 AU cavity (dashed lines). The models with depletion factors of $10^3$, $10^4$ and $10^5$ are indicated with green, blue and black lines respectively.}
    \label{fig:NH2}
\end{figure}

\begin{figure*}[t]
    \includegraphics[scale=0.4]{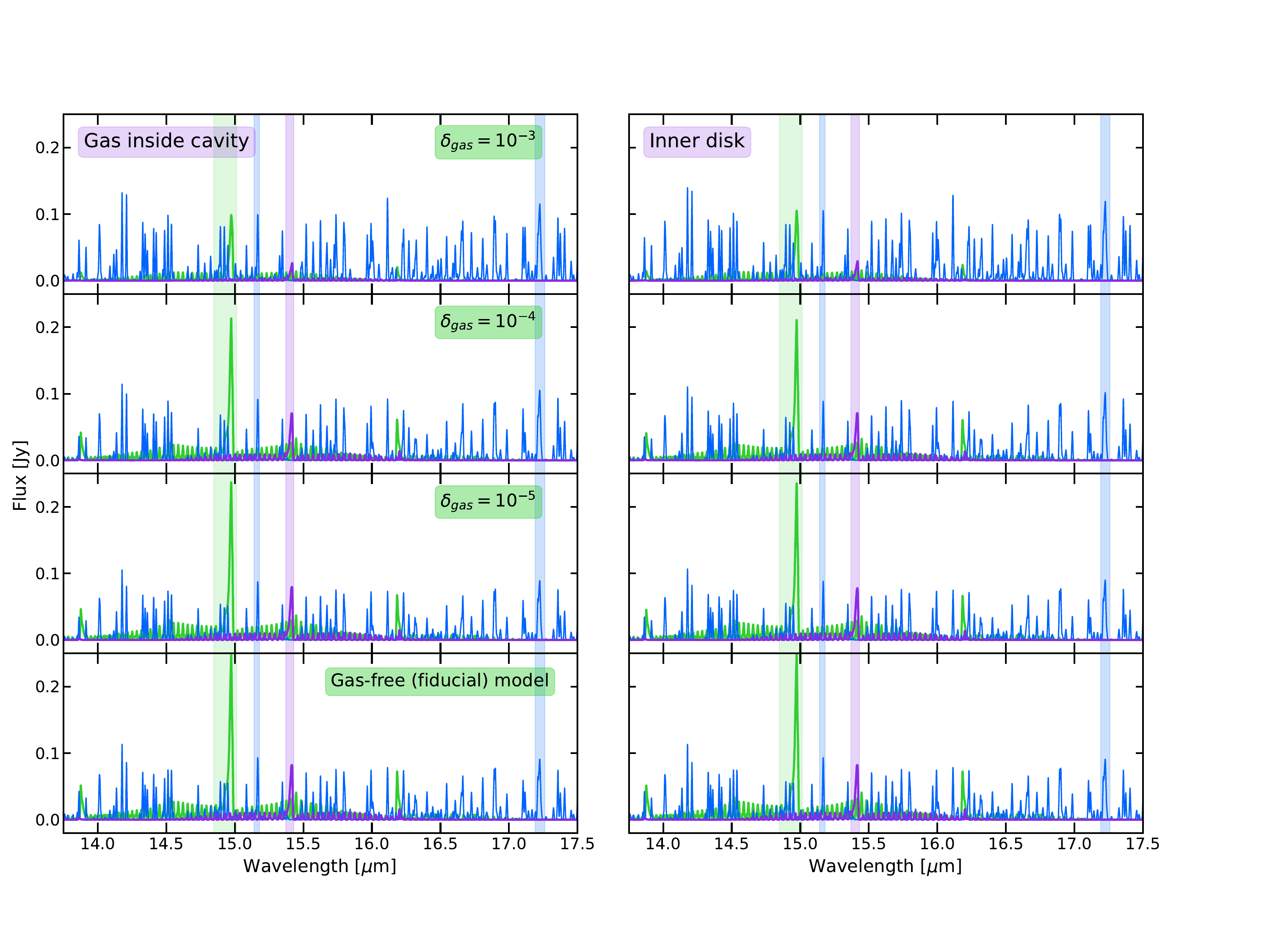}
    \caption{Generated \ce{H2O}, \ce{^12CO2} and \ce{^13CO2} spectra for the models with gas inside a 5 AU cavity, but no inner disk (left column) or an 0.5 AU inner disk inside a 5 AU, completely empty cavity (right column). The gas and dust inside the cavity or the inner disk are depleted by factors of $10^3$, $10^4$ and $10^5$ (first, second and third rows respectively). The fourth row shows, in both panels, the fiducial model with a 5 AU cavity, where the cavity is completely empty. The vertical colored bars indicate the integration ranges for the \ce{H2O} 15.17 and 17.22 $\mu$m flux (blue), as well as the \ce{^12CO2} and \ce{^13CO2} Q-branches (green and purple respectively).}
    \label{fig:spectra_dgas_innerdisk}
\end{figure*}
\subsection{Alternative scenarios} \label{subsec:altexpls}

So far, we have tested the effects of a deep cavity in the gas and dust on the \ce{H2O} and \ce{CO2} spectra of a T Tauri disk, and we have also tested the influence of the stellar luminosity on this scenario. Naturally, there are many other parameters or variations to be made that could be interesting for further study. Here we discuss three of them, and we present a few additional models to provide a quick insight into these scenarios, though we leave a thorough investigation for future work.

\subsubsection{Gas and dust inside the cavity}
All of the models presented here so far consider a cavity that is fully depleted in both gas and dust. However, this is not necessarily representative for all disks, since their stars are often still accreting at a decent rate. It is possible for some amount of gas and dust to still be flowing through the cavity, or there may still be a small inner disk present. Observations of many transitional disks show that gas is present inside large dust cavities \citep[see, e.g.][]{vandermarel2015, vandermarel2016}, at various levels of depletion, and small inner disks have been found inside the large cavities as well \citep{francis2020}.\\
\newline
Naturally, if the cavity is not fully depleted in gas, this may contribute to the spectrum. To understand how this could impact our results, we examine each of these two scenarios with six additional models. All of these models contain a 5 AU cavity, since this model shows a spectrum that is dominated by a bright \ce{CO2} Q-branch in our fiducial grid. As such, this will help us quantify how much gas and dust inside the cavity is needed to impact our results. We create three models in which some gas and dust are present over the full extent of the cavity, depleted by factors of $10^3$, $10^4$ and $10^5$ with respect to the outer disk. We also create three models in which a small inner disk is present out to 0.5 AU, and the rest of the cavity is empty. This inner disk is then also depleted in gas and dust by the same factors with respect to the outer disk. \\
\newline
In Fig. \ref{fig:NH2}, we show the \ce{H2} column density as a function of disk radius for these six additional models. The models with an 0.5 AU inner disk are indicated using solid lines and the models with gas inside the entire extent of the cavity are indicated using dashed lines. To understand what level of gas depletion in the inner disk could be considered realistic, we can compare our models to the work of \citet{leemker2022}. They find an upper limit on the total hydrogen column density $N_{\rm H} \sim 10^{22}$ cm$^{-2}$ inside the cavity of LkCa15, consistent with the finding of $\sim 10^{21}$ cm$^{-2}$ from \citet{salyk2009}. Looking at studies from, for example, \citet{salyk2009} and \citet{brown2013}, who derive the CO column density in the inner disk, it is seen that total hydrogen columns of $10^{21-23}$ cm$^{-2}$ are common. This corresponds best to our models with depletion factors of $10^4$ and $10^5$.\\
\begin{figure*}[t]
    \includegraphics[scale=0.38]{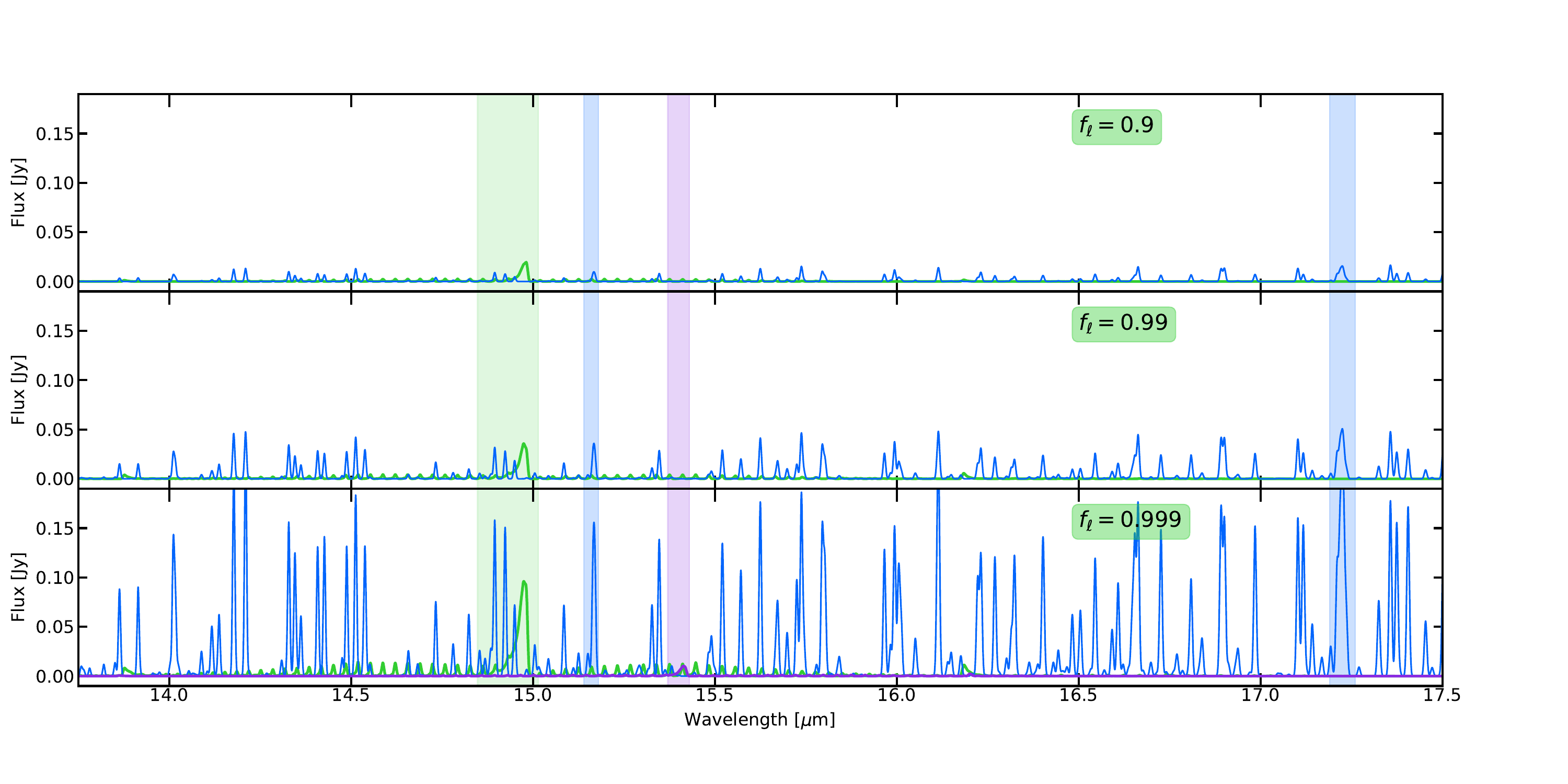}
    \caption{Generated \ce{H2O}, \ce{^12CO2} and \ce{^13CO2} spectra for the models with $f_{\ell} = 0.9, 0.99$ and 0.999 (the fiducial model). All are full disk models with no cavity present. The vertical colored bars indicate the integration ranges for the \ce{H2O} 15.17 and 17.22 $\mu$m flux (blue), as well as the \ce{^12CO2} and \ce{^13CO2} Q-branches (green and purple respectively).}
    \label{fig:spectra_flarge}
\end{figure*}
\newline
The spectra generated for these six models, compared with the fiducial model with a 5 AU cavity, are presented in Fig. \ref{fig:spectra_dgas_innerdisk}. The three models with gas and dust inside the entire cavity are (including the fiducial model) are shown in the left column, and the three models with an 0.5 AU inner disk (plus fiducial model) are shown in the right column. This figure shows that, if enough gas and dust is present inside the cavity, this can indeed start to impact our results. For depletion factors of $10^3$ and less, the gas inside the cavity contributes significantly to the spectrum. This enhances the \ce{H2O} emission w.r.t. the \ce{CO2}, since this gas is hot enough to contain abundant \ce{H2O}. Depletion factors larger than $10^4$ are needed for the contribution from the gas inside the cavity to be negligible and for most of the emission to originate from outside of the cavity. So, we can see from these tests that the cavity needs to be sufficiently depleted of gas and dust for our previous results to hold. The necessary depletion factors are consistent with a typical amount of gas in the inner disk, following e.g. \citet{salyk2009, brown2013, leemker2022}.\\  
\newline
The gas and dust are depleted by equal factors in the models previously shown, but one could also envision a scenario in which the cavity contains significantly more gas than dust (for example due to the dust being locked up in larger bodies). In this case, the gas temperature inside the cavity will rise significantly \citep[see, e.g.][]{leemker2022}, making the molecular emission brighter. We do not expect this to significantly affect the \ce{CO2 /H2O} ratio, however, since the emission from both molecules will likely be boosted by roughly equal amounts.

\subsubsection{Effect of the gas-to-dust ratio}
\begin{figure}[t]
    \includegraphics[width=\linewidth]{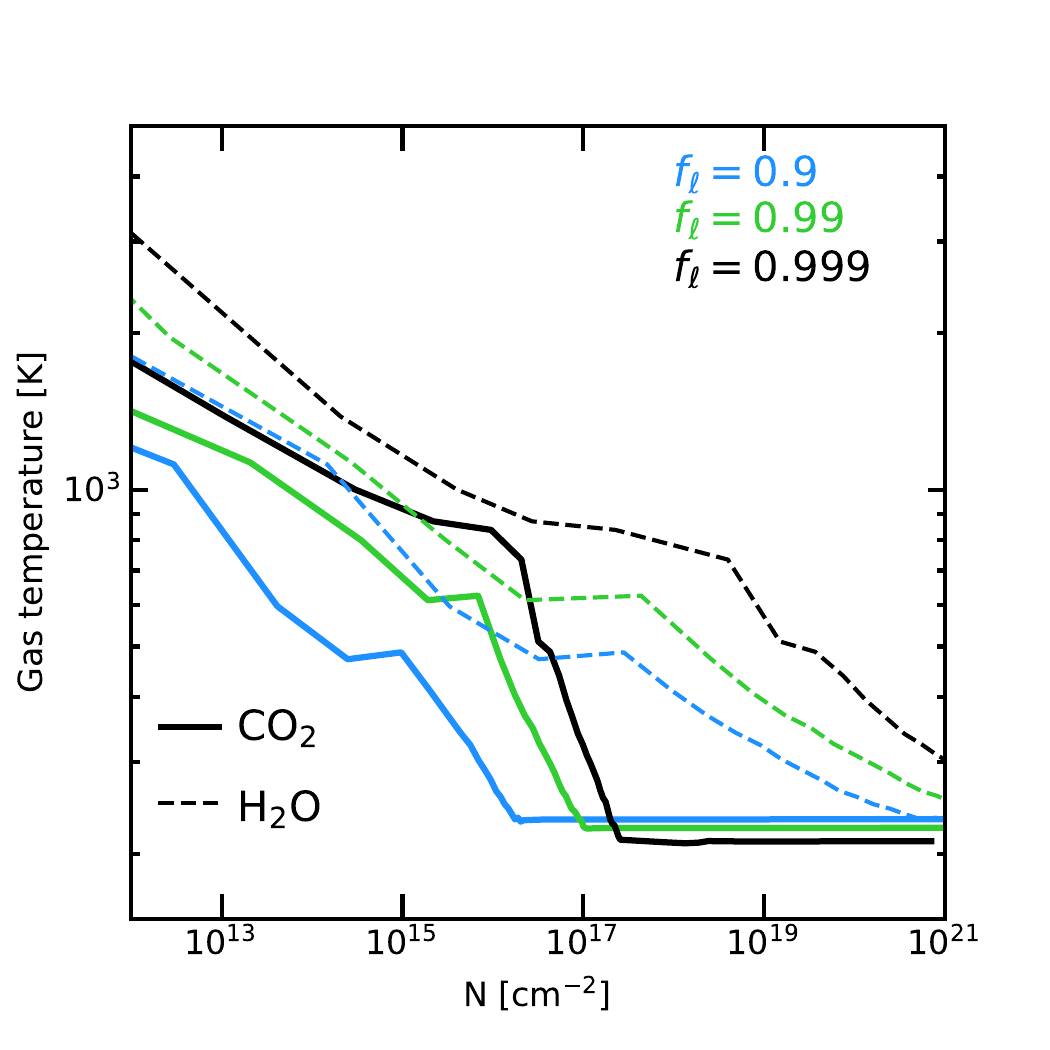}
    \caption{Gas temperature in a vertical cut at 0.3 AU, plotted as a function of \ce{CO2} (solid lines) and \ce{H2O} (dashed lines) column density. This is shown for the full disk models with $f_{\ell} = 0.9$ (blue), 0.99 (green) and 0.999 (black; the fiducial model).}
    \label{fig:flarge_N_tgas}
\end{figure}

It is well known that the amount of small grains in the upper layers of the disk can also have a significant effect on the observed mid-IR fluxes.
It was shown by \citet{meijerink2009} that mid-IR \ce{H2O} lines observed with \textit{Spitzer} likely originate from a region where the gas-to-dust ratio is locally increased by 1-2 orders of magnitude compared to the canonical interstellar medium (ISM) value of 100. This is necessary to match the observed line strengths and line-to-continuum ratios of typical sources. We incorporate this into our models by assuming an overall gas-to-dust mass ratio of 100, but we assume that a significant fraction (99.9\%) of the dust is large and is settled to the midplane. This leaves the disk atmosphere relatively depleted in dust, and this area will have a gas-to-dust ratio of $10^5$ instead. This yields large line fluxes and high line-to-continuum ratios, which strongly decrease as the amount of dust in the upper layers of the disk increases \citep[see, e.g.][]{woitke2018, greenwood2019}. The \ce{CO2}-only sources often have weak \ce{H2O} fluxes, possibly pointing to smaller gas-to-dust ratios in these sources. The \ce{CO2 /H2O} flux ratio may also be impacted by the gas-to-dust ratio, as a larger amount of dust in the disk atmosphere will cool this region more, possibly stimulating the gas-phase formation of \ce{CO2} over the formation of \ce{H2O}. \\
\newline
To provide a preliminary test to this scenario, we present two additional models (based on our fiducial grid) in which we vary the fraction of large grains, $f_{\ell}$, and thus the gas-to-dust ratio in the upper layers of the disk (see also Sect. \ref{sec:methods}). The two additional models are full disk models, so they do not have a cavity, and they have $f_\ell = 0.9$ and 0.99 (corresponding to gas-to-dust ratios of $10^3$ and $10^4$ in the upper layers, respectively). The spectra for these models are presented in Fig. \ref{fig:spectra_flarge}. This shows that a lower gas-to-dust ratio in the upper layers ($f_{\ell} = 0.9$) can indeed produce a \ce{CO2}-dominated spectrum, completely without the need to introduce a cavity into the model. However, one can also see that the absolute flux levels drop significantly \citep[see also][]{bosman2017, woitke2018}. This could provide a diagnostic to test whether a source is more likely to be \ce{CO2}-dominated due to a small cavity or due to a small gas-to-dust ratio, since the drop in \ce{CO2} flux could become too large to be consistent with the latter.\\
\newline
It is likely that the relative evolution of \ce{CO2} and \ce{H2O} line emission with gas-to-dust ratio is an effect of both temperature and chemistry. The effect of the gas-to-dust ratio on the gas temperature is presented in Fig. \ref{fig:flarge_N_tgas}, where the gas temperature is plotted as a function of the \ce{CO2} and \ce{H2O} column densities (solid and dashed lines, respectively) for all three models. This figure clearly shows a decrease in gas temperature in the upper layers of the disk as $f_{\ell}$, and thus the gas-to-dust ratio, decreases. Since \ce{CO2} is preferentially formed over \ce{H2O} at lower temperatures, this could explain the shift in flux ratio. The lower temperature could also be a cause of the lower absolute flux of both molecules.  \\
\newline
Large gas-to-dust ratios in the upper layers of the disk are needed to reproduce mid-IR fluxes \citep{meijerink2009, bosman2022a}, so that may point to a cavity being a better explanation. Still, this scenario is very interesting to consider, especially in relation to disks like GW Lup and CX Tau, which (as we discuss in Sect. \ref{subsec:comp_obs}) show even higher \ce{CO2 /H2O} flux ratios than this work predicts, yet all of the observational evidence collected so far does not necessarily point to a large-enough cavity being present in these disks. Another interesting example to consider here is IM Lup, a known \ce{CO2}-only disk as observed by \textit{Spitzer} \citep{pontoppidan2010, bosman2017}. \citet{bosman2023} speculate that this disk could have a high abundance of small grains in its upper layers due to strong radial drift and vertical mixing. The authors speculate that this could hide the disks \ce{H2O} emission behind optically thick dust, yet pumping by IR continuum photons could boost other emission features such as the \ce{CO2} Q-branch, providing an alternative explanation as to why this disk is \ce{CO2}-bright. More in-depth testing is needed to confirm this theory, though it further highlights that a more thorough exploration of the gas-to-dust ratio and how it impacts mid-IR molecular emission is warranted. This will be done in future work. 

\subsubsection{A dust trap instead of a cavity}
One final alternative scenario that will be qualitatively considered here is that of a dust trap with only a small perturbation in the gas surface density. If a dust trap were located between the \ce{H2O} and \ce{CO2} snowlines, this could prevent the flow of \ce{H2O}-ice-rich pebbles from reaching the disk region inside the \ce{H2O} snowline, preventing the enhancement of the \ce{H2O} gas that is formed in-situ by sublimation of ices \citep[e.g.][]{banzatti2017, banzatti2020}. This could lead to the inner regions of the disk becoming relatively depleted in \ce{H2O} gas, especially if the gas in the inner disk (in which \ce{H2O} can still form in the gas-phase) is accreted onto the star at a rapid rate. This would not be true for the \ce{CO2} gas, since this could be replenished by sublimation of ices. This higher abundance of \ce{CO2} compared to \ce{H2O} in the inner disk could then possibly explain the relatively stronger \ce{CO2} emission. \\
\newline
Dust traps and the radial drift of ices can also have some interesting effects on the C/O ratio in the inner regions. Modeling work from \citet{mah2023} shows that, even without a dust trap present, the C/O ratio in the inner regions of the disk will increase over time due to the inward advection of C-rich gas from beyond the \ce{H2O} snowline. They also show that this process is significantly faster for low-mass stars, potentially explaining their high observed C/O ratios. Similarly, a dust trap could trap O-rich ices outside the \ce{H2O} snowline, raising the C/O ratio in the inner regions. If this O-depletion is significant enough, the emission from the inner disk may get a significant contribution from hydrocarbons. The emission from both \ce{H2O} and \ce{CO2} would likely be strongly reduced in such a case. Since \ce{CO2} contains two O atoms, and would thus be doubly affected by the O-depletion caused by the dust trap, the dust trap would likely need to be located sufficiently close to the \ce{H2O} snowline for the emission of \ce{CO2} to be stronger than that of \ce{H2O}. DALI does not take pebble transport into account, but such a scenario could still be modeled with the code by varying the initial abundances of C and O, and thus the C/O ratio, with disk radius. Such an exploration is also left for future work.

\section{Conclusions} \label{sec:conclusions}
Some sources observed with \textit{Spitzer} and JWST show a very high \ce{CO2 /H2O} flux ratio. In this work, we present DALI thermo-chemical models to investigate one scenario which may cause this: the presence of a small, inner cavity in the disk that extends beyond the \ce{H2O} snowline, but not beyond the \ce{CO2} snowline. We also test for the effects of stellar luminosity as well as \ce{H2O} ro-vibrational cooling. Our main results are summarized below.

\begin{enumerate}
    \item 
    A spectrum that is \ce{H2O}-dominated when no cavity is present in the disk can become \ce{CO2}-dominated by introducing a large-enough cavity into the disk. 
    \item 
    The cavity size needed to achieve this is roughly 4-5 AU for our fiducial, $L_* = 1.4\,L_\odot$ model (vs. 2-3 AU for the $L_* = 0.2 \, L_\odot$ model) and is found to be consistently larger than the nominal midplane \ce{CO2} snowline location in a full disk, contrary to our initial hypothesis. As such, the presence of this small, inner cavity could more easily be tested observationally. 
    \item 
    The inclusion of \ce{H2O} ro-vibrational transitions in DALI's heating and cooling balance creates a small decrease in gas temperature of at most a factor of two in the layer from which \ce{H2O} emits, which in turn lowers the \ce{H2O} line fluxes by roughly the same amount. In our models, this local gas temperature change results in the spectrum becoming \ce{CO2}-dominated whereas it would be \ce{H2O}-dominated were these transitions not included.
    \item 
    Global temperature changes, such as those caused by a difference in stellar luminosity, do not seem to affect the relative \ce{CO2 /H2O} flux ratio much, instead impacting both molecules' fluxes equally.
    \item 
    Comparison with JWST-MIRI data of four T Tauri disks from the MINDS GTO program finds two disks which are clearly \ce{H2O}-dominated (DR Tau and Sz 98) and two which are clearly \ce{CO2}-dominated (GW Lup and CX Tau), the latter of which may have a small, inner cavity. The observational evidence for this is so far inconclusive.
    \item 
    If enough gas and dust is still present inside the cavity, this could boost the \ce{H2O} lines w.r.t. the \ce{CO2}. However, for typical inner disk hydrogen column densities, this effect is negligible. 
    \item 
    A lower gas-to-dust ratio can also lead to a spectrum becoming \ce{CO2}-dominated, without the presence of a cavity, though this also causes the absolute line fluxes to drop steeply. A more thorough exploration of this and other possible scenarios will be conducted in future work. 
\end{enumerate}

\begin{acknowledgements}
    We thank Arthur Bosman for the insightful discussions, Ardjan Sturm for comments on the manuscript, and the MINDS GTO team for providing the MIRI data shown in this work.
    Astrochemistry in Leiden is supported by funding from the European Research Council (ERC) under the European Union’s Horizon 2020 research and innovation programme (grant agreement No. 101019751 MOLDISK).  B.T. is a Laureate of the Paris Region fellowship program, which is supported by the Ile-de-France Region and has received funding under the Horizon 2020 innovation framework program and Marie Sklodowska-Curie grant agreement No. 945298.
\end{acknowledgements}

%
%
\bibliographystyle{aa}
\bibliography{references}

\begin{appendix}
\section{Including \texorpdfstring{\ce{H2O}}{H2O} rovibrational cooling}
\label{sec:app_rovib}

We created an additional grid of models to test the effects of \ce{H2O} ro-vibrational cooling. As described in Sect. \ref{sec:methods}, in the second step of DALI's model creation, the excitation balance of several specified atoms and molecules is calculated. For this purpose, a file specifying all of the molecule's transitions is supplied. On the Leiden Atomic and Molecular Database, two versions of such a file are available for \ce{H2O}, one including the molecule's ro-vibrational and pure rotational transitions, and one containing only its pure rotational transitions. For the calculation of the fiducial grid (as well as the lower-luminosity grids), only the pure rotational transitions of \ce{H2O} are considered in DALI's heating and cooling balance, as is also done in the work by \citet{bosman2022a, bosman2022b}, on which the fiducial grid is based. For the additional grid presented in this section, the ro-vibrational transitions of \ce{H2O} are also included. The inclusion of \ce{H2O} ro-vibrational cooling is the only change made to the models for this additional grid, all other parameters are kept the same. As is shown in Sect. \ref{subsec:rovib}, this inclusion causes a small, localized decrease in the gas temperature, which in turn translates into a decrease in the \ce{H2O} line fluxes. All models presented in this work also contain cooling through CO ro-vibrational transitions (vibrational levels $\nu \leq 2$ included, collisional rate coefficients from \citealt{yang2010}) and OH pure rotational transitions \citep[collisional rate coefficients from][]{offer1994}. Cooling through \ce{CO2} emission lines is not considered in any of the models. 

\begin{figure*}[t]
    \makebox[\textwidth][c]{\includegraphics[scale=0.5]{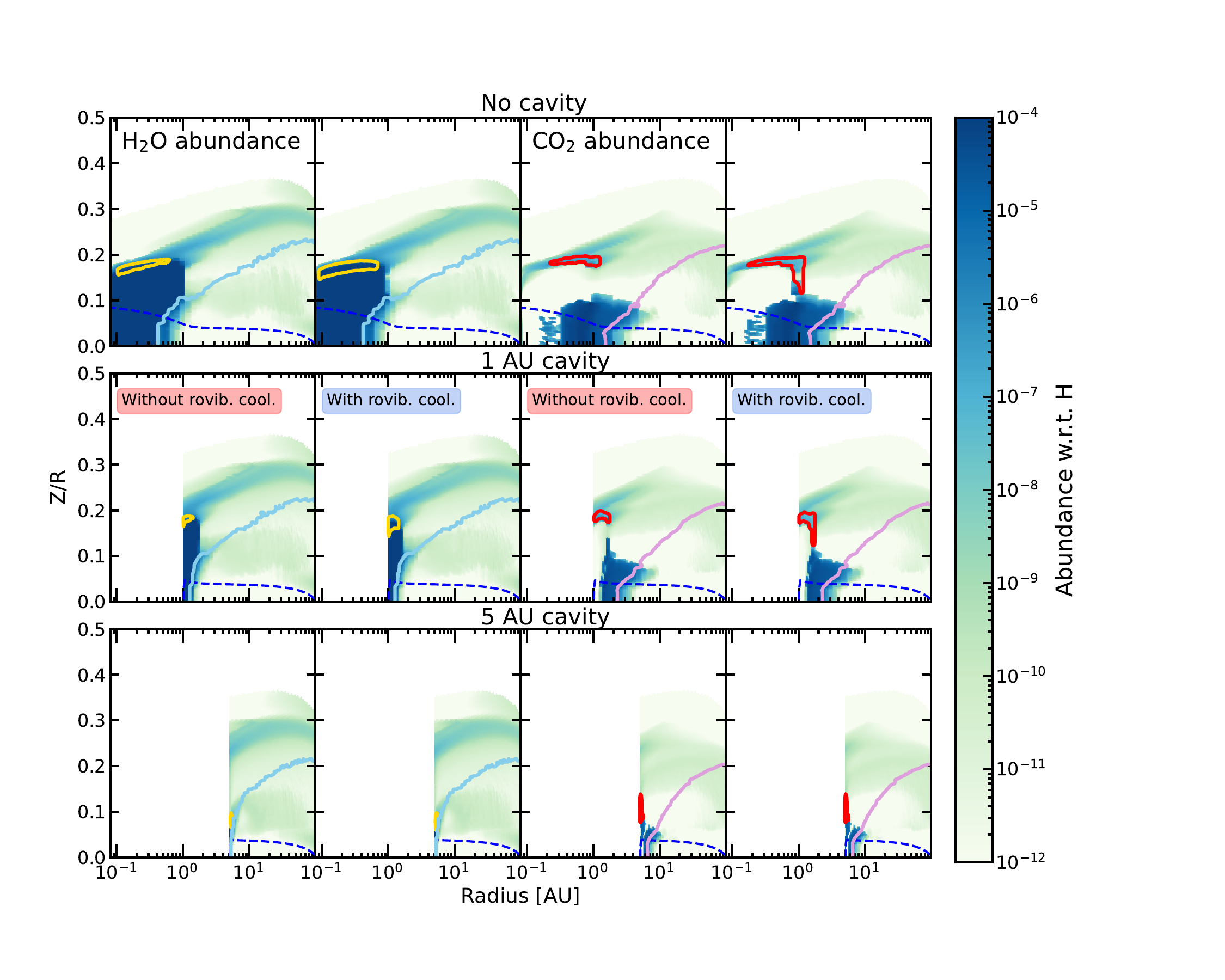}}
    \caption{Abundance maps of \ce{H2O} (first and second columns) and \ce{CO2} (third and fourth columns) for the models with no cavity (top row), a 1 AU cavity (middle row) and a 5 AU cavity (bottom row) for the fiducial grid without \ce{H2O} ro-vibrational cooling (first and third columns) and the grid with \ce{H2O} ro-vibrational cooling (second and fourth columns). The \ce{H2O} and \ce{CO2} snowlines (defined as $n_{\rm gas}/n_{\rm ice} = 1$) are indicated with solid light blue and pink lines respectively. The dashed, dark blue line shows the dust $\tau=1$ surface at 15 $\mu$m. The yellow contours indicate the region in which 90\% of the \ce{H2O} $11_{3,9} - 10_{0,10}$ line emission at 17.22 $\mu$m ($E_{\rm up} = 2438$ K) originates. The red contours indicate the origin of 90\% of the \ce{CO2} $01^10 - 00^00$ Q(20) line emission ($E_{\rm up} = 1196$ K).}
    \label{fig:h2o_co2_rv_nonrv}
\end{figure*}

\begin{figure*}[t]
    \makebox[\textwidth][c]{\includegraphics[scale=0.5]{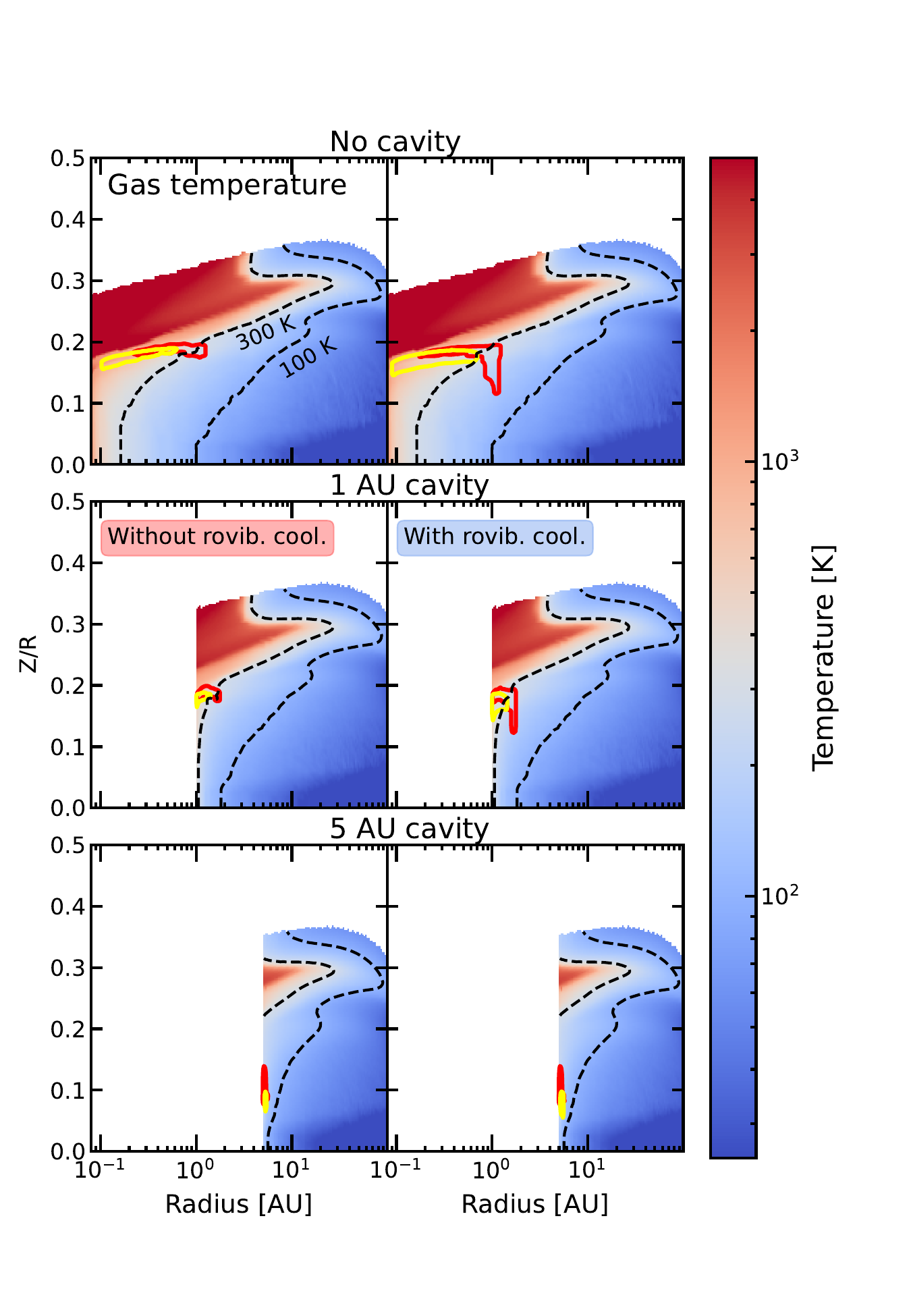}}
    \caption{Gas temperature map for the models with no cavity (top row), a 1 AU cavity (middle row) and a 5 AU cavity (bottom row) for the fiducial grid without \ce{H2O} ro-vibrational cooling (left column) and the grid with \ce{H2O} ro-vibrational cooling (right column). Temperature contours of 100 and 300 K are indicated with dashed, black lines. The yellow contours indicate the region in which 90\% of the \ce{H2O} $11_{3,9} - 10_{0,10}$ line emission at 17.22 $\mu$m ($E_{\rm up} = 2438$ K) originates. The red contours indicate the origin of 90\% of the \ce{CO2} $01^10 - 00^00$ Q(20) line emission ($E_{\rm up} = 1196$ K).}
    \label{fig:tgas_rv_nonrv}
\end{figure*}

\begin{figure*}[t]
    \makebox[\textwidth][c]{\includegraphics[scale=0.3]{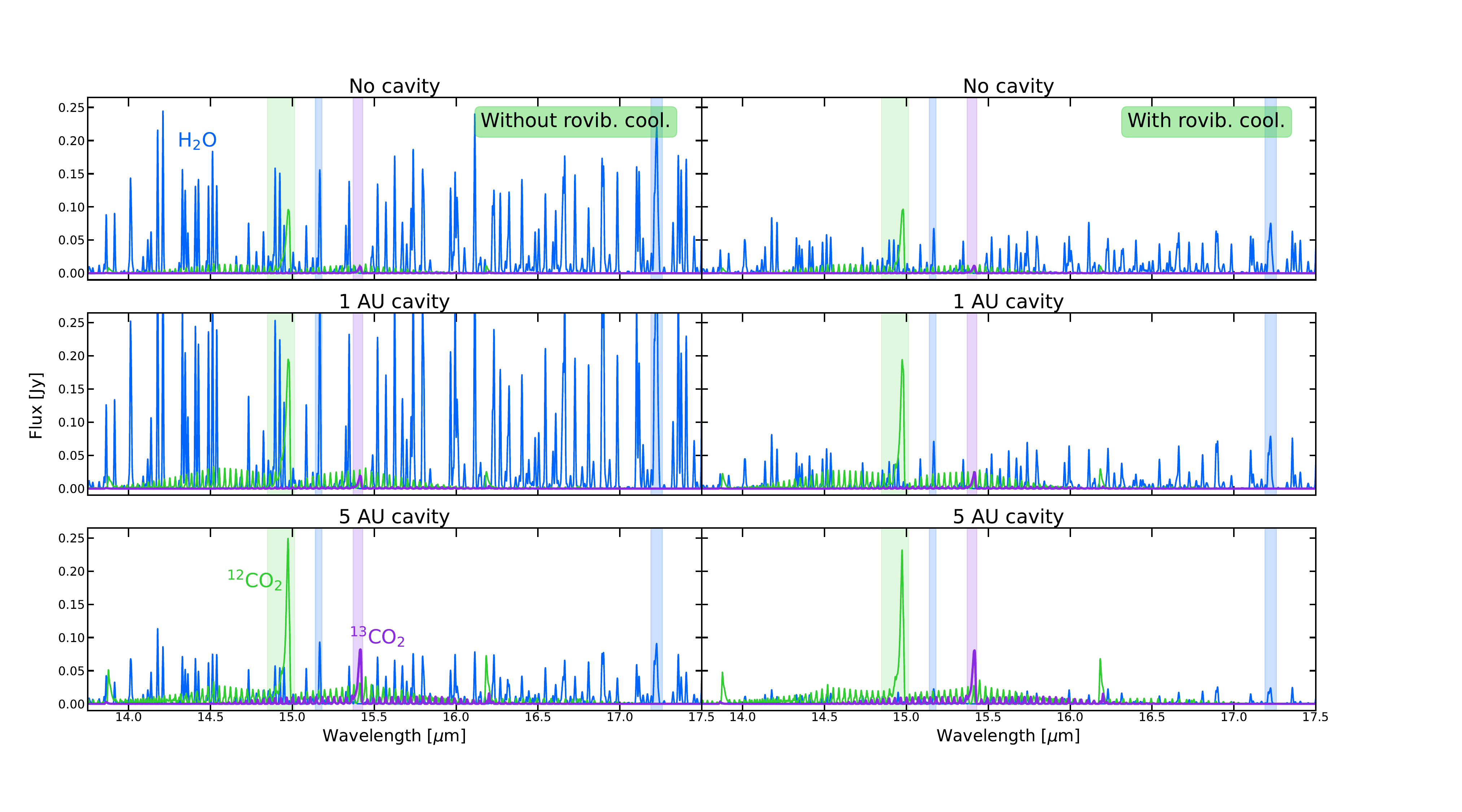}}
    \caption{Generated \ce{H2O}, \ce{^12CO2} and \ce{^13CO2} spectra for the fiducial grid without \ce{H2O} ro-vibrational cooling (left column) and the grid with \ce{H2O} ro-vibrational cooling (right column). The spectra are shown for the models with no cavity (top row), a 1 AU cavity (middle row) and a 5 AU cavity (bottom row). The vertical colored bars indicate the integration ranges for the \ce{H2O} 15.17 and 17.22 $\mu$m flux (blue), as well as the \ce{^12CO2} and \ce{^13CO2} Q-branches (green and purple respectively).}
    \label{fig:spectra_subset}
\end{figure*}

\begin{figure*}[ht]
    \makebox[\textwidth][c]{\includegraphics[scale=0.4]{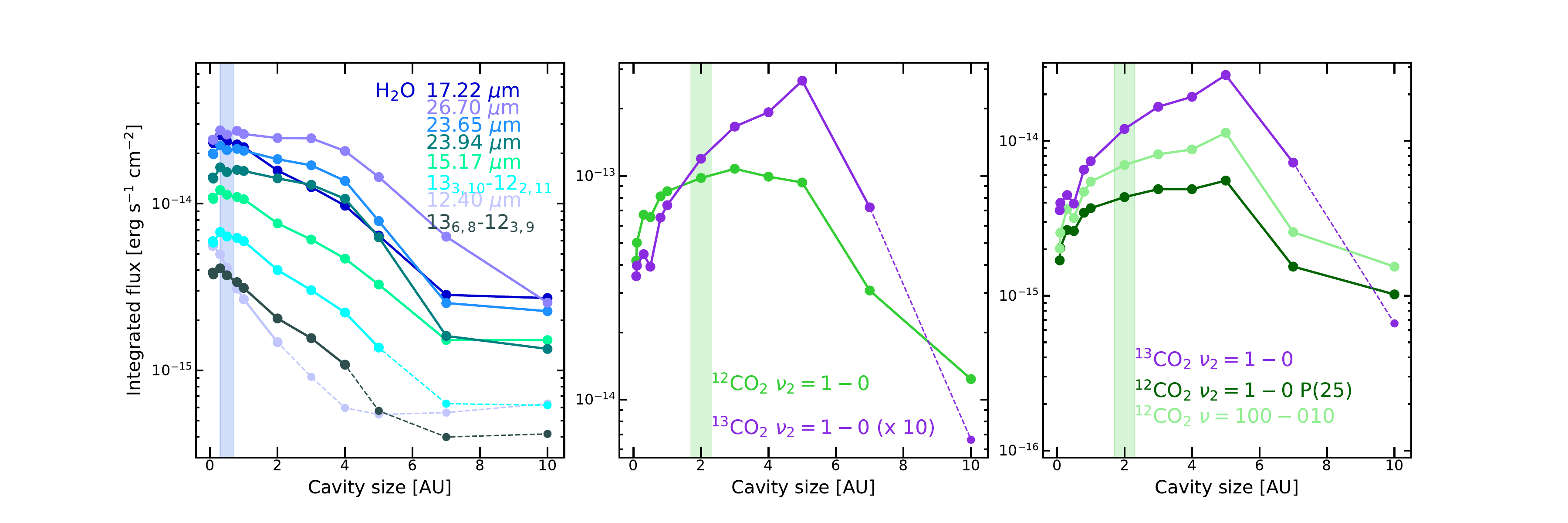}}
    \caption{Integrated fluxes of several \ce{H2O} (left panel),  \ce{^12CO2} and \ce{^13CO2} (middle and right panels) line complexes plotted as a function of cavity size for grid with \ce{H2O} ro-vibrational cooling. Properties of these line complexes can be found in Table \ref{tab:Eup}. Nominal \ce{H2O} and \ce{CO2} midplane snowlines in the full disk model are indicated with colored bars. Data points are connected by dashed lines if the line flux drops below $10^{-15}$ erg s$^{-1}$ cm$^{-2}$, indicating that the lines will become more difficult to observe.}
    \label{fig:abs_flux_rv}
\end{figure*}

\begin{figure}[ht]
    \includegraphics[scale=0.45]{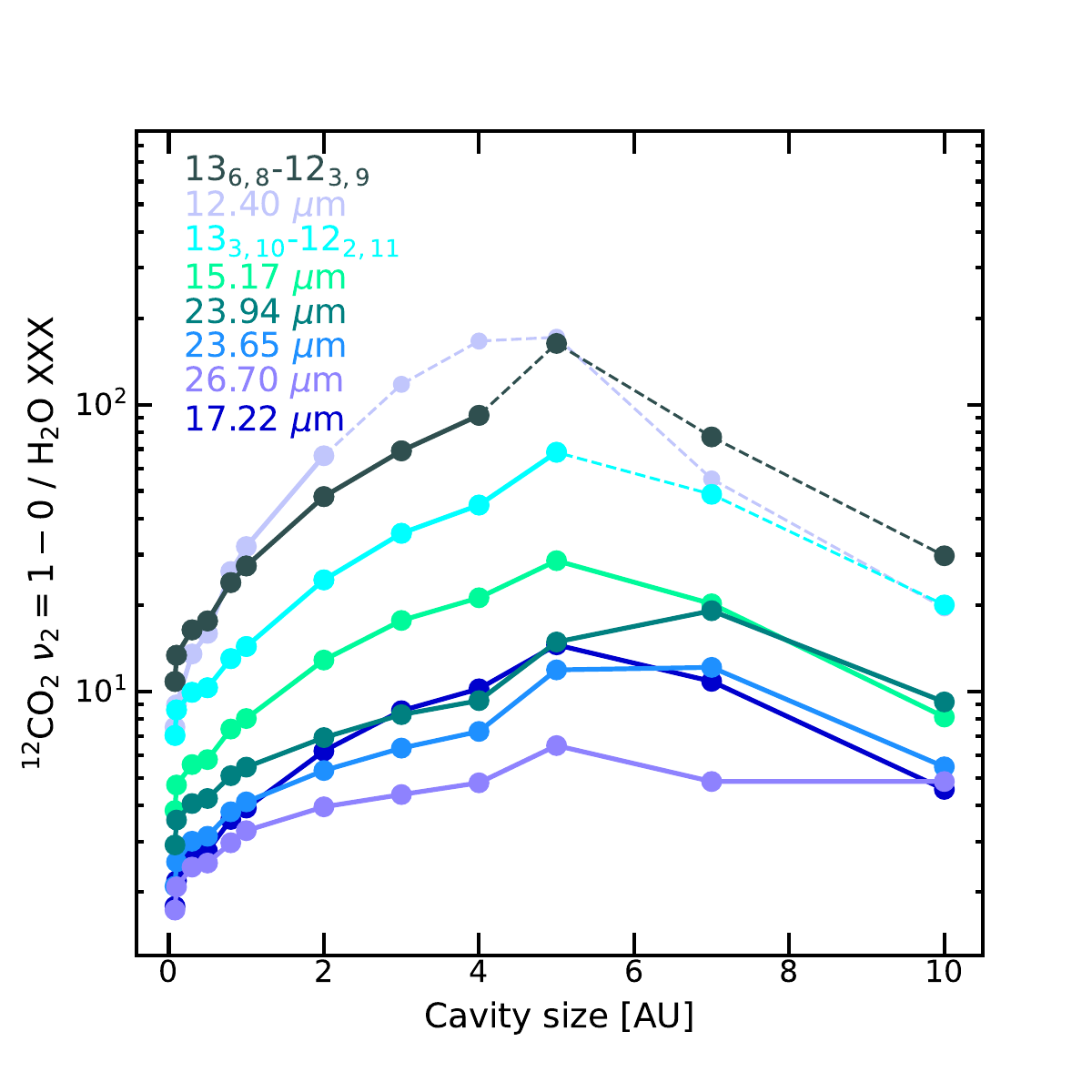}
    \caption{Ratio of the \ce{^12CO2} Q-branch flux and the fluxes of several \ce{H2O} line complexes, plotted as a function of cavity size for the grid with \ce{H2O} ro-vibrational cooling. Properties of these line complexes can be found in Table \ref{tab:Eup}. Data points are connected by dashed lines if the line flux drops below $10^{-15}$ erg s$^{-1}$ cm$^{-2}$, indicating that the lines will become more difficult to observe.}
    \label{fig:rel_flux_rv}
\end{figure}

\section{Lower-luminosity grids}
\label{sec:app_lum}

As described in Sect. \ref{sec:intro}, the stellar luminosity may affect the \ce{H2O} and \ce{CO2} spectra of a disk as well, with disks around lower-luminosity stars being colder and thus potentially having a more \ce{CO2}-dominated spectrum. Hence, we performed our analysis not just for our fiducial grid of models based on AS 209, but also for three grids with central stars with luminosities of 0.2, 0.4 and 0.8 $L_\odot$. For comparison, the ``\ce{CO2}-only'' source GW Lup observed with JWST by \citet{grant2023} has $L_* = 0.3\, L_\odot$. For these grids, the input spectra consist of a blackbody at temperatures of 3500, 3750 and 4000 K respectively, all with an additional 10000 K component modeling the UV luminosity generated by accretion \citep{kama2016, visser2018}. The lower luminosity input spectra are generated to have far-ultraviolet (FUV; 911--2067 \AA) luminosities of $\log L_{\rm FUV}/L_\odot = -3.45, -2.92$ and $-2.62$ for the $L = 0.2, 0.4$ and $0.8\,L_\odot$ grids respectively. This corresponds to a mass accretion rate of $\log \dot{M}_{\rm acc} = -9.5$ for the lowest luminosity spectrum, whereas for the two higher-luminosity spectra this corresponds to a mass accretion rate of $\log \dot{M}_{\rm acc} = -9$. The dust sublimation radius is located at 0.03, 0.04 and 0.06 AU for the $L_* = 0.2$, 0.4, and 0.8 $L_\odot$ grids respectively. 

\begin{figure*}[t]
    \makebox[\textwidth][c]{\includegraphics[scale=0.5]{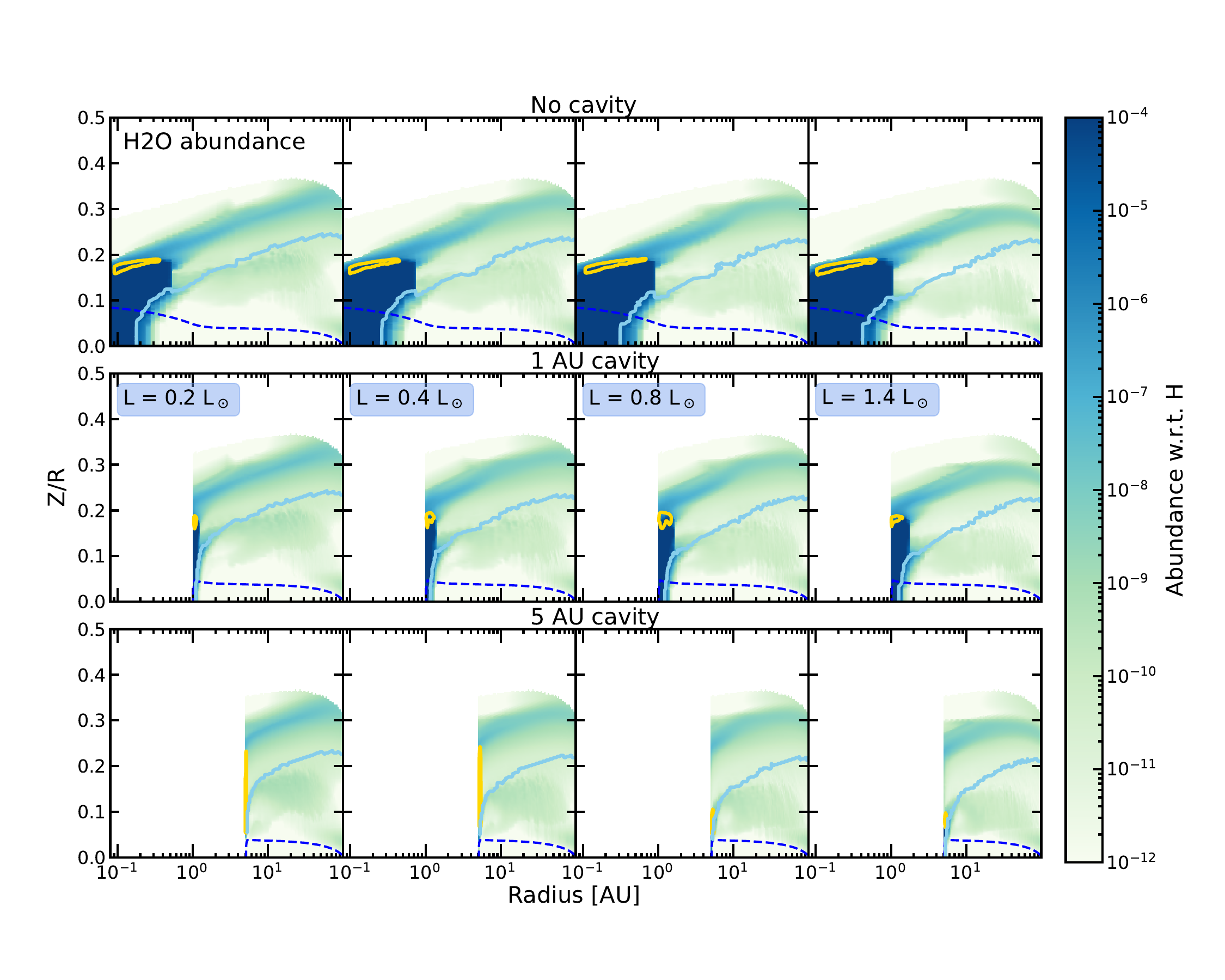}}
    \caption{Abundance maps of \ce{H2O} for the models with no cavity (top row), a 1 AU cavity (middle row) and a 5 AU cavity (bottom row) for the grids with an input stellar spectrum with $L_* = 0.2\,L_\odot$ (first column), $L_* = 0.4\,L_\odot$ (second column), $L_* = 0.8\,L_\odot$ (third column), and $L_* = 1.4\,L_\odot$ (the fiducial grid, fourth column). The \ce{H2O} snowline (defined as $n_{\rm gas}/n_{\rm ice} = 1$) is indicated with a solid light blue line. The dashed, dark blue line shows the dust $\tau=1$ surface at 15 $\mu$m. The yellow contours indicate the region in which 90\% of the \ce{H2O} $11_{3,9} - 10_{0,10}$ line emission at 17.22 $\mu$m ($E_{\rm up} = 2438$ K) originates.}
    \label{fig:h2o_comp_lowL}
\end{figure*}

\begin{figure*}[t]
    \makebox[\textwidth][c]{\includegraphics[scale=0.5]{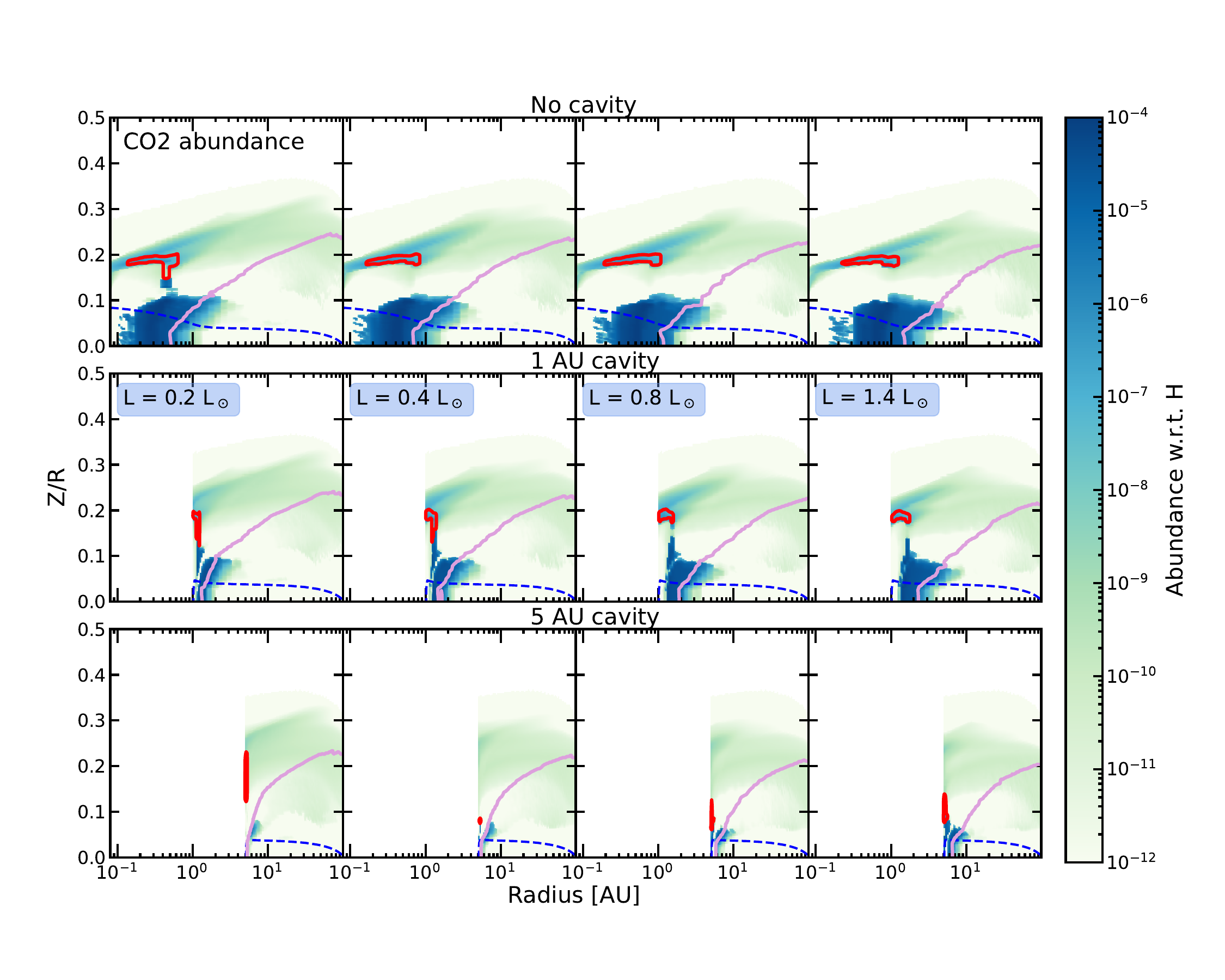}}
    \caption{Abundance maps of \ce{CO2} for the models with no cavity (top row), a 1 AU cavity (middle row) and a 5 AU cavity (bottom row) for the grids with an input stellar spectrum with $L_* = 0.2\,L_\odot$ (first column), $L_* = 0.4\,L_\odot$ (second column), $L_* = 0.8\,L_\odot$ (third column), and $L_* = 1.4\,L_\odot$ (the fiducial grid, fourth column). The \ce{CO2} snowline (defined as $n_{\rm gas}/n_{\rm ice} = 1$) is indicated with a solid light blue line. The dashed, dark blue line shows the dust $\tau=1$ surface at 15 $\mu$m. The red contours indicate the origin of 90\% of the \ce{CO2} $01^10 - 00^00$ Q(20) line emission ($E_{\rm up} = 1196$ K).}
    \label{fig:co2_comp_lowL}
\end{figure*}

\begin{figure*}[t]
    \makebox[\textwidth][c]{\includegraphics[scale=0.5]{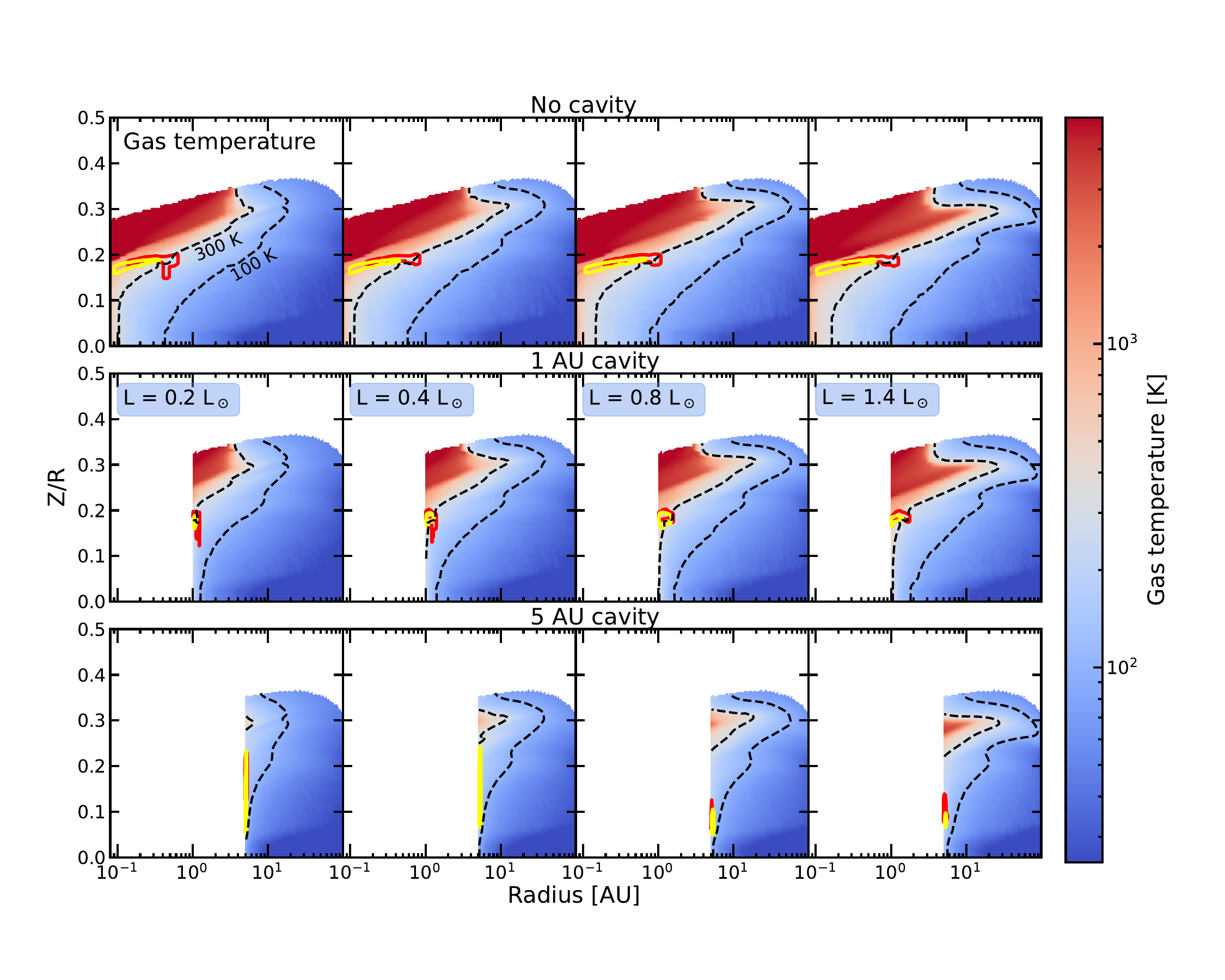}}
    \caption{Gas temperature maps for the models with no cavity (top row), a 1 AU cavity (middle row) and a 5 AU cavity (bottom row) for the grids with an input stellar spectrum with $L_* = 0.2\,L_\odot$ (first column), $L_* = 0.4\,L_\odot$ (second column), $L_* = 0.8\,L_\odot$ (third column), and $L_* = 1.4\,L_\odot$ (the fiducial grid, fourth column). Temperature contours of 100 and 300 K are indicated with dashed black lines. The yellow contours indicate the region in which 90\% of the \ce{H2O} $11_{3,9} - 10_{0,10}$ line emission at 17.22 $\mu$m ($E_{\rm up} = 2438$ K) originates. The red contours indicate the origin of 90\% of the \ce{CO2} $01^10 - 00^00$ Q(20) line emission ($E_{\rm up} = 1196$ K).}
    \label{fig:tgas_comp_lowL}
\end{figure*}

\begin{figure*}[t]
    \makebox[\textwidth][c]{\includegraphics[scale=0.28]{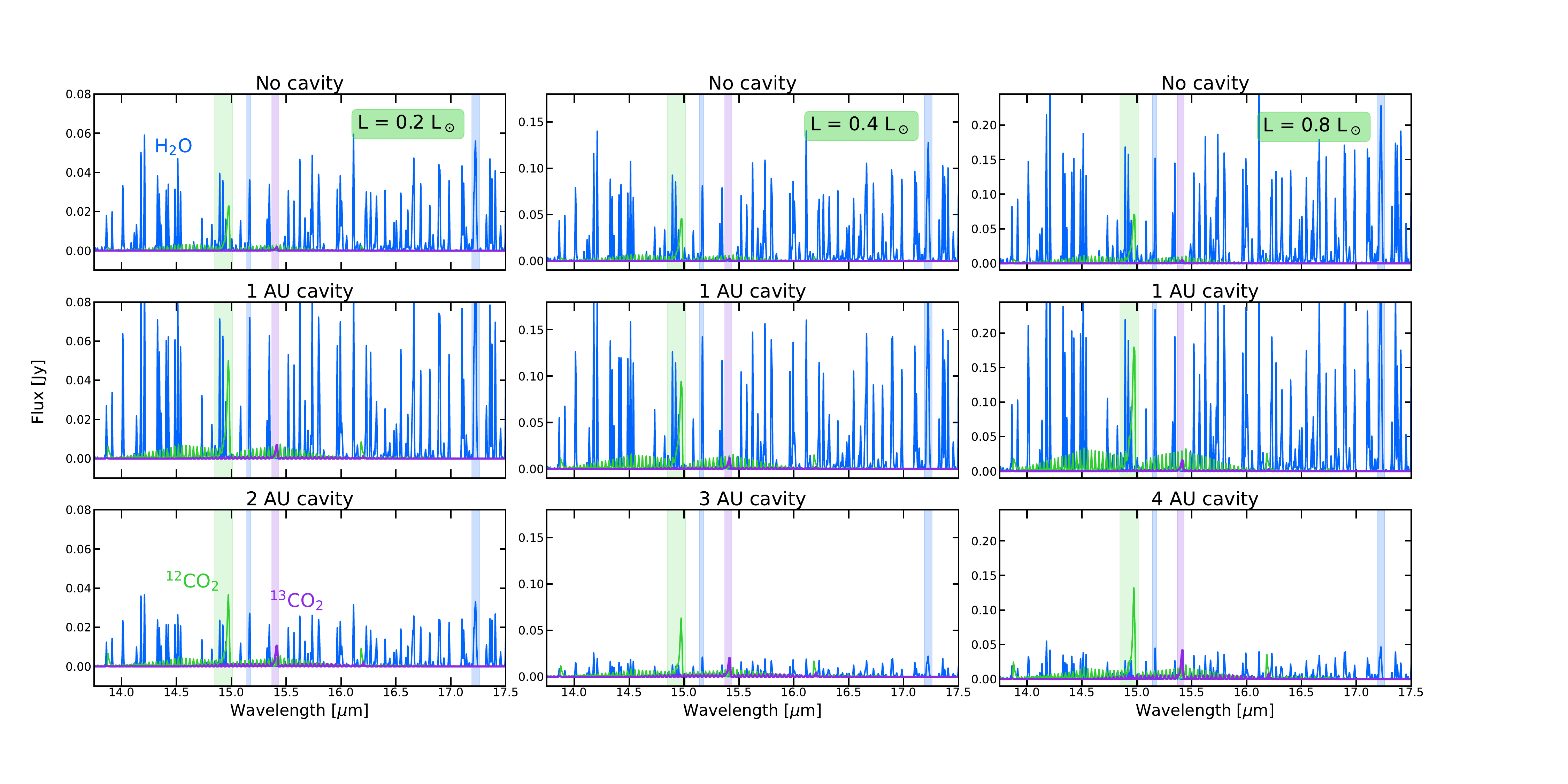}}
    \caption{Generated \ce{H2O}, \ce{^12CO2} and \ce{^13CO2} spectra for the grids with $L_* = 0.2, 0.4$ and $0.8\,L_\odot$. The spectra are shown for the models with no cavity (top row), a 1 AU cavity (middle row) and a 2 AU, 3 AU or 4 AU cavity (cavity size at which the spectrum becomes \ce{CO2}-dominated, bottom row, left, middle and right panels respectively). The vertical colored bars indicate the integration ranges for the \ce{H2O} 15.17 and 17.22 $\mu$m flux (blue), as well as the \ce{^12CO2} and \ce{^13CO2} Q-branches (green and purple respectively). Note that the scaling on the y-axis is different for each column.}
    \label{fig:spectra_subset_lowL}
\end{figure*}

\begin{figure*}[ht]
    \makebox[\textwidth][c]{\includegraphics[scale=0.5]{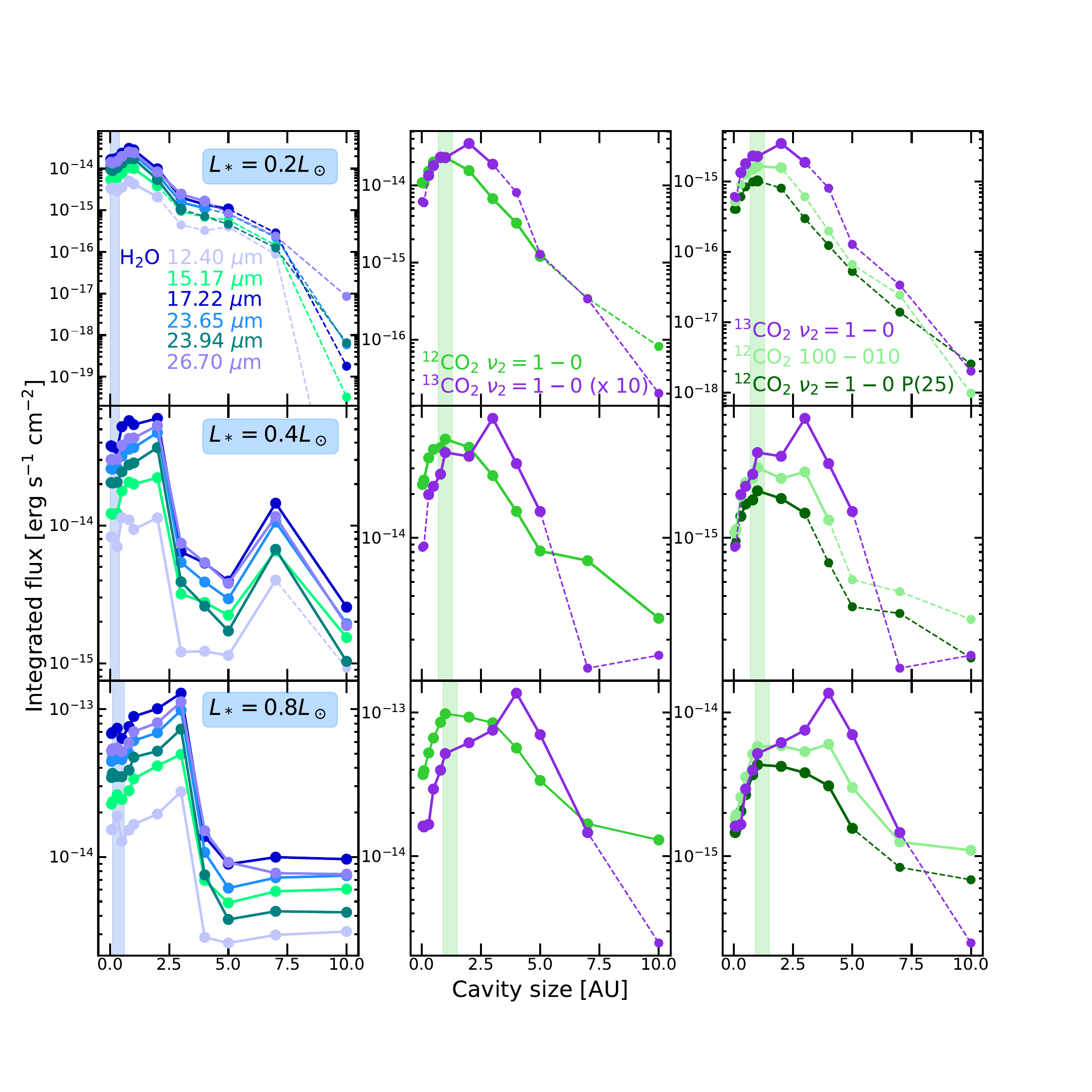}}
    \caption{Integrated fluxes of several \ce{H2O} (left panel),  \ce{^12CO2} and \ce{^13CO2} (middle and right panels) line complexes plotted as a function of cavity size for the grids with input spectra with $L_* = 0.2, 0.4, 0.8\,L_\odot$. Properties of these line complexes can be found in Table \ref{tab:Eup}. Nominal \ce{H2O} and \ce{CO2} midplane snowlines in the full disk model are indicated with colored bars. Data points are connected by dashed lines if the line flux drops below $10^{-15}$ erg s$^{-1}$ cm$^{-2}$, indicating that the lines will become more difficult to observe.}
    \label{fig:abs_flux_lowL}
\end{figure*}

\begin{figure*}[ht]
    \makebox[\textwidth][c]{\includegraphics[scale=0.4]{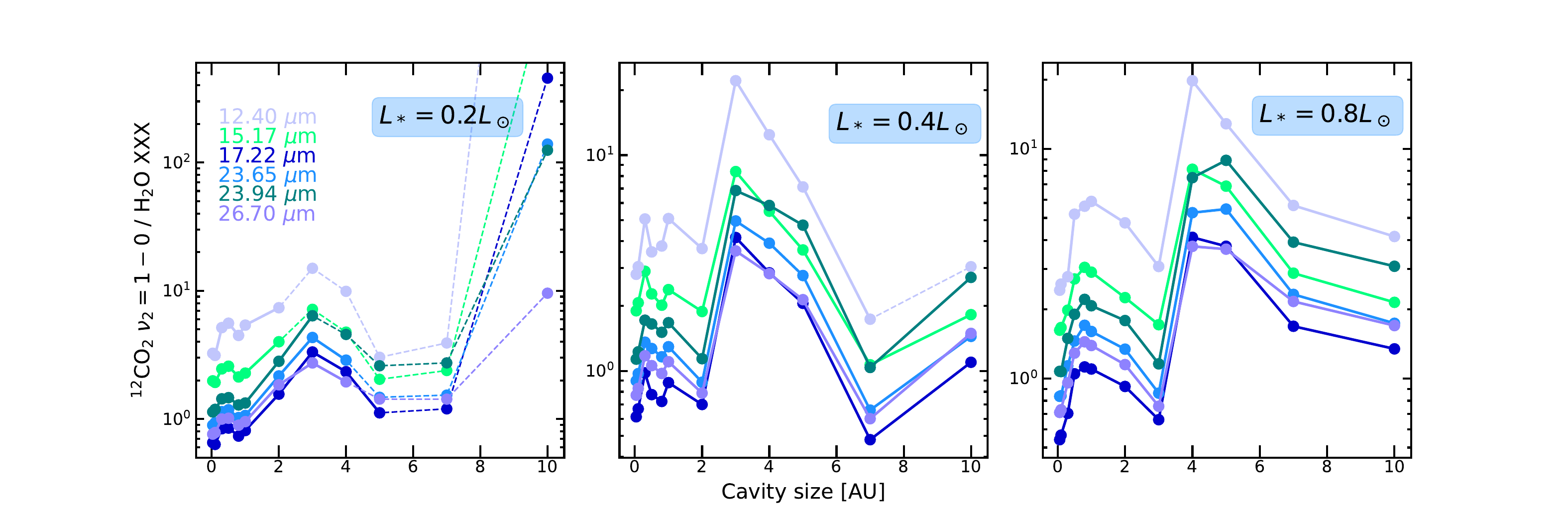}}
    \caption{Ratio of the \ce{^12CO2} Q-branch flux and the fluxes of several \ce{H2O} line complexes, plotted as a function of cavity size for the models with input spectra with $L_* = 0.2, 0.4, 0.8\,L_\odot$. Note the different vertical scales of these panels. Properties of these lines can be found in Table \ref{tab:Eup}. Data points are connected by dashed lines if the line flux drops below $10^{-15}$ erg s$^{-1}$ cm$^{-2}$, indicating that the lines will become more difficult to observe.}
    \label{fig:rel_flux_lowL}
\end{figure*}

\begin{figure}[t]
    \includegraphics[scale=0.45]{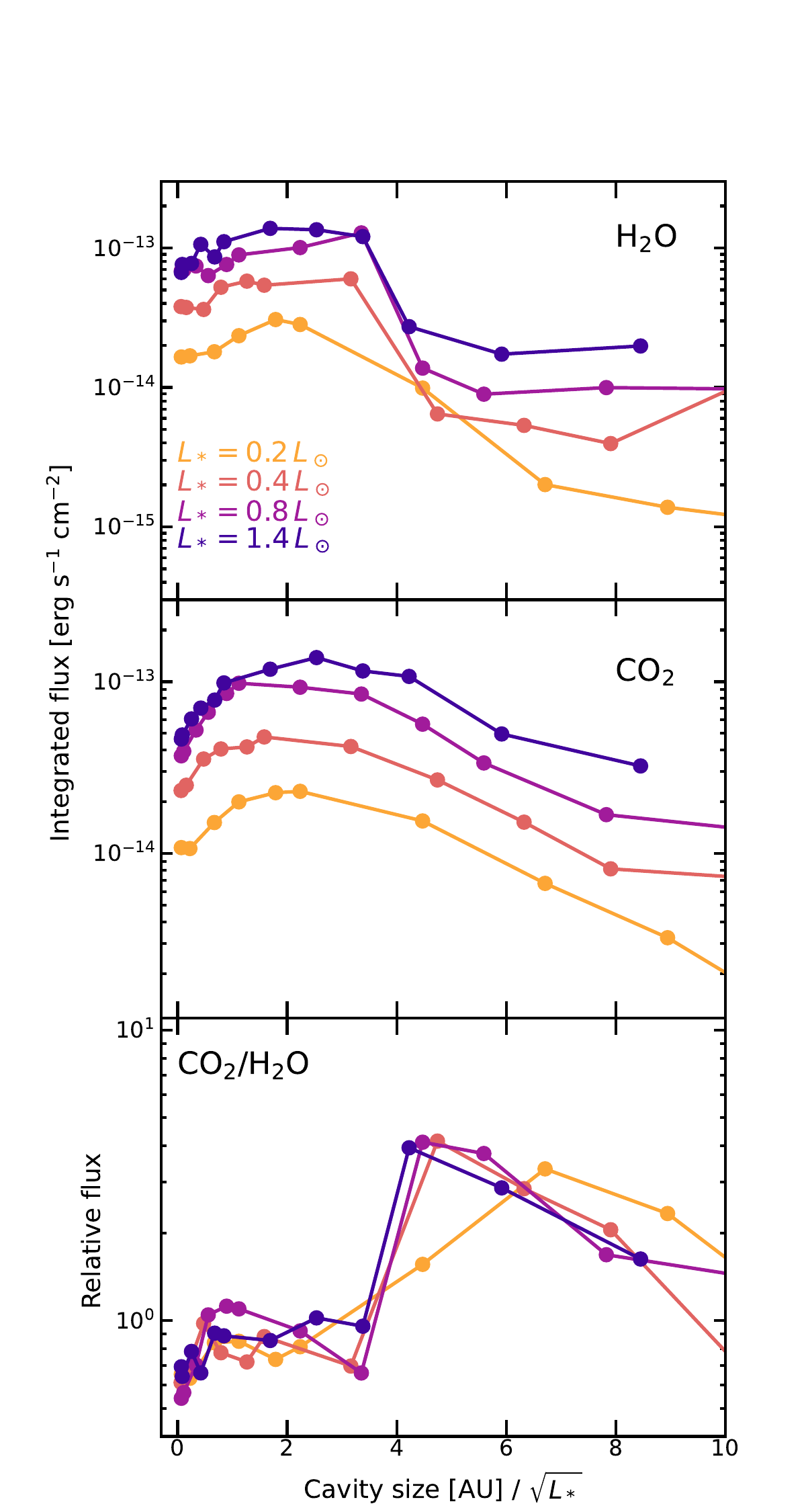}
    \caption{Flux of the 17.22 $\mu$m \ce{H2O} line complex (top), the \ce{CO2} Q-branch (middle) and their ratio (bottom) for the grids with $L_* = 0.2, 0.4, 0.8$ and $1.4\,L_\odot$, plotted as a function of cavity size scaled with $\sqrt{L_*}$.}
    \label{fig:flux_lowL_rescaled}
\end{figure}

\end{appendix}

\end{document}